\documentclass{llncs}
\usepackage{fullpage}
\usepackage{amsmath, amssymb, graphicx, url,
    cite, subfigure, color}

\newcommand{\remove}[1]{}

\begin{document}

\title{Towards Plugging Privacy Leaks in Domain Name System}
\author{{\rm Yanbin Lu and Gene Tsudik}}
\institute{Computer Science Department, University of California, Irvine\\
\{yanbinl, gene.tsudik\}@uci.edu\\ }
\maketitle

\begin{abstract} 
Privacy leaks are an unfortunate and an integral part of the current
Internet domain name resolution. Each DNS query generated by a user
reveals -- to one or more DNS servers -- the origin and target of
that query. Over time, a user's browsing behavior might be exposed
to entities with little or no trust. Current DNS privacy leaks stem
from fundamental features of DNS and are not easily fixable by simple
patches. Moreover, privacy issues have been overlooked by DNS
security efforts (i.e. DNSSEC) and are thus likely
to propagate into future versions of DNS.

In order to mitigate privacy issues in current DNS, this paper
proposes a Privacy-Preserving Domain Name System (PPDNS), which
maintains privacy during domain name resolution. PPDNS is based on
distributed hash tables (DHTs), an alternative naming infrastructure,
and computational private information retrieval (cPIR)
\cite{Gentry2005SingleDatabase}, an advanced cryptographic
construct. PPDNS takes advantage of the DHT's index structure to
improve name resolution query privacy, while leveraging cPIR to reduce
communication overhead for bandwidth-sensitive clients. Our analysis
shows that PPDNS is a viable approach for obtaining a higher degree of
privacy for name resolution queries. PPDNS also serves as a
demonstration of blending advanced systems techniques with their
cryptographic counterparts.
\end{abstract}

\section{Introduction}
Domain Name System (DNS) \cite{rfc1034} is a global
hierarchically distributed database. It translates human-readable
hostnames into numerical identifiers associated with network-layer
IP interface addresses for the purpose of locating (and routing to)
individual hosts. Given the importance of DNS, much effort has been put 
into the next generation of secure DNS, referred to as DNSSEC
\cite{rfc4033}.  The main purpose of DNSSEC is to authenticate the
origin, and guarantee the integrity of, DNS records. Notably, DNSSEC
explicitly rules out data disclosure threats \cite{rfc3833}.

It is easy to see that today's DNS provides no privacy. All DNS messages are
transmitted in the clear. A malicious local name server that 
delegates DNS queries for all clients in its administrative domain
can easily gain all origin-target information. It also knows each
target's query volume from its administrative domain. Each name server
in the DNS zone hierarchy -- from the root to the authoritative server
for a target queried by a local name server on behalf of its client --
learns both the target of the query and query volume for the
target. Although caching at the local name server hides the exact
volume from other name servers in the hierarchy, authoritative servers
can still infer expected query volume based on the distribution of
queries' arrival. Different leaks of DNS query pose differnt risks.


Leaks of query target's identity can be abused for censorship
purposes. For example, an ISP can forbid access to a host by simply
dropping all queries for that host-name at the local name server
controlled by that ISP. This approach, known as DNS filtering, is
effectively employed by Great Firewall of China~\cite{greatfirewall}
to filter out websites considered as a threat by the state. Also it is
part of the upcoming German national ISP filtering plan and similar
efforts by Australian and New Zealand governments are already
partially deployed. Although this is not the only way for an ISP to
block a host, it is more effective than other approaches.  First, a
host name is easier memorizable than an IP address; hence, most people
prefer to use host-names.  Second, IP blocking may not work if a host
can have a large pool of IP addresses and frequently change the
mapping between its host-name and an IP address unknown to the ISP.
If there was a perfect scheme for privately resolving host-names, the
only conclusion the ISP would draw is that there is traffic between
the client and an unknown IP address\footnote{Assuming IPSec
  \cite{rfc4033} can be used to hide all information above IP
  layer.}. Of course, the ISP can use reverse DNS to retrieve the
host-name corresponding to the unknown IP address. However, a host can
simply configure its PTR record to some random name to prevent the ISP
from learning anything useful.  Another approach ISP can use is to
install a proxy bound to all sensitive host names, hoping to catch all
traffic directed to those host names. However, DNSSEC can effectively
prevent this from happening.

Leaks of target query volume can be used for surveillance purposes.
For example, if a DNS server observes a sudden spike in the number of
DNS queries for a given host-name, it can alert authorities to check
whether suspicious or offensive information on that host is causing
its increased popularity. This is a common surveillance skill used in
countries such as China to check possible sensitive websites.

Leaks of relative target query volume (which maybe inferred from
target query volume) can be abused for commercial purposes. For
example, suppose that the adversary can measure the relative
popularity of any two domain names registered under a certain
registrar. Then, the adversary might steal a popular domain name when
its owner fails to renew it. One may think the chance is quite low for a
popular domain owner to forget to register its domain name.  However,
in history, even industry giants like Microsoft once forgets to renew
its domain names~\cite{microsoft_forget}. Many registrars have been
making it more difficult for such theft to occur with auto-renewals,
e.g., by parking domains for a grace period.  However, if the
adversary happens to be the registrar itself, it can reserve any popular
domains registered under it and sell them at higher prices later.

Moreover, leaks of information during DNS query also pose risk to
applications which rely on DNS as an underlying mechanism. For
example, ENUM~\cite{rfc3761} (Telephone number mapping) system uses
special DNS record types to translate a traditional telephone number
into a Uniform Resource Identifier from VoIP providers, as well as to
other Internet-based services such as E-mail and web page. The exposure
of DNS query means the revealing of one's daily contact list, which is
considered to be a serious privacy betrayal.


This paper focuses on adversaries at DNS servers. It is true that a
malicious DNS server can pose more threats than those mentioned above.
We limit the scope of threats here based on the assumption that a
powerful adversary does not want to expose itself too early by
performing too aggressively. Although the above-mentioned leaks of
information maybe collected by adversaries through other means, for
adversaries at DNS servers, learning that information through DNS is
a straightforward way. This paper focuses on addressing target-related
DNS query privacy problems by proposing a privacy-preserving domain
name system (PPDNS). PPDNS combines the use of a distributed hash
table (DHT) with a computational private information retrieval (cPIR)
\cite{Gentry2005SingleDatabase} technique. Owing to consistent
identifiers offered by DHT, the same range query can be posed for each
identifier inside the range even in the presence of an active
adversary, thereby hiding the actual query target.  Moreover, we can
also achieve the same range query for each identifier inside the range
among all clients, thereby hiding the volume and relative volume for a
query target. Furthermore, each DHT node can treat each range as a
small database and take advantage of cPIR which allows a client to
retrieve a record from a server 
(without revealing which record is being retrieved) with less than
$\mathcal{O}(N)$ communication overhead; where $N$ is the number of
database records.

The contribution of this work is as follows: (1) we investigate and
categorize DNS privacy leaks, (2) we propose a novel range query technique
that mitigates privacy leaks, (3) we design a flexible and
parallelizable framework for incorporating cPIR at the server side,
and (4) we provide a security analysis showing that PPDNS offers
appreciably better privacy than prior work.

The rest of this paper is organized as follows:
Sec.~\ref{sec:legacy_overview} provides an overview of current DNS.
Sec.~\ref{sec:adv_model} formally defines the adversary model.
Sec.~\ref{sec:related} analyzes some existing approaches and
explains why they are insecure. Sec.~\ref{sec:blocks} introduces
some preliminaries before presenting PPDNS.  The architecture of
PPDNS is given in Sec.~\ref{sec:architecture}. Detailed security
analysis comes in Sec.~\ref{sec:analysis}.
Sec.~\ref{sec:performance_eval} shows the performance of PPDNS and
Sec.~\ref{sec:conclusion} concludes the paper.

\section{DNS Overview \label{sec:legacy_overview}}
\begin{figure}[!t]
\centering
\includegraphics[width=3.2in]{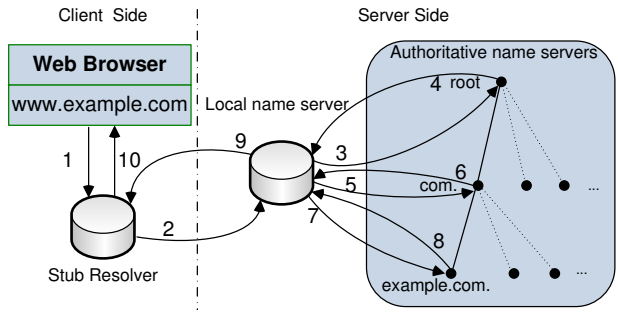}
\caption{\small  Current DNS Infrastructure:
 ten-step resolution of domain name ``www.example.com''.}
 \label{fig:dns_illus}
\end{figure}
The DNS~\cite{rfc1034, rfc1035} namespace has a tree structure where
each node, except the root, has a non-empty label. Each domain is a
node in the namespace tree, and bottom-up concatenation of nodes'
labels delimited by periods (``."), creates a fully qualified domain
name. For instance, \texttt{www.example.com} is a fully qualified
DNS name of a node \texttt{www} with the parent  \texttt{example},
grandparent -- \texttt{com}, and  great-grandparent -- DNS root.

Nodes are grouped together into zones. The apex of a zone is called
the {\em start of authority} (SOA) and bottom edges are called {\em
  delegation points } if other zones exist below them, or {\em leaf
  nodes}, otherwise. Zones are served by \textit{authoritative name
  servers} that are either {\em primary}, if the zone data comes to
them from the outside of DNS, or {\em secondary}, if their zone data
comes to them from primary servers via a zone transfer
procedure. Authoritative name servers manage all name information in
the domain, keep track of authoritative name servers of their
sub-domains rooted at their domain, and are administered by namespace
operators.

Every node can store {\em resource records} (RRs) containing information
associated with a domain name. A RR can map a host name to an IP
address or vice versa, or serve a variety of other purposes. Each RR
has a name, class, type, TTL and data. A {\em RRset} is a list of all the
records matching a given domain name and resource type. A nameserver
returns a RRset as a response to a query.

DNS initiators on host machines are called \textit{stub resolvers};
they do not have caches of their own and do not directly interact
with the zone hierarchy. They pose basic queries to
\textit{local name servers}, also known as recursive resolvers,
within their own administrative domain. The local name server is
usually designated by the ISP and only provides service to hosts
within its administrative domain. It accepts
recursive queries from stub resolvers, contacts a chain of
authoritative name servers inside the zone hierarchy to locate a DNS
RRset, and answers the stub resolver. Because pursuing a chain of
delegations to resolve a query can incur significant delays, the local
name server caches the result and answers subsequent queries using the
cache, until a TTL (assigned by the authoritative nameserver) expires.

In an example shown in Fig. \ref{fig:dns_illus} (which assumes no
caching), the resolution of \texttt{www.example.com} involves
following steps: (1) The user types the domain name into the web
browser which consults its local stub resolver, usually implemented as
a set of library routines. (2) The stub resolver sends the query to
its local name server. (3) The local name server consults with the
root server -- the authority for the empty label. (4) The root server
refers the local name server to ``com.'' zone's authoritative
server. (5) The local name server consults a ``com.'' zone's
authoritative server. (6) It delegates the local name server to
``example.com'' zone's authoritative server.  (7) The local name
server consults ``example.com'' zone authoritative server. (8) This
authoritative server answers the host address record for
``www.example.com''. (9) Finally the local name server caches the
resource record and meanwhile returns the result to the stub
resolver. (10) The stub resolver returns the IP address to the web
browser which then launches connection to the IP address.

\section{Threat Analysis \label{sec:adv_model}}
In this section we discuss DNS privacy threats and the adversarial model.

\subsection{DNS Privacy Leaks \label{sec:privacy_leak}}
We consider the following DNS privacy leaks/problems:
\begin{enumerate}
\item query source-target association
\item query source identity
\item source query volume, i.e., number of DNS queries issued by a given source.
\item relative source query volume, i.e., the difference between query volumes for two given sources.
\item query target identity
\item target query volume, i.e., number of DNS queries issued for a given target.
\item relative target query volume, i.e., the difference between query volumes for two given targets.
\end{enumerate}
Of course, DNS clients may choose to use a general-purpose anonymization service, such as
TOR~\cite{tor} to hide the DNS query source. This prevents problems
(1)-(4). However, source anonymity afforded by tools like TOR does nothing to address
target-specific leaks (5)-(7). In the rest of this paper, we focus on the latter.

\subsection{Adversary Model\label{sec:adv_type}}
An adversary can be an insider or an outsider. Outsider is an
adversary that does not compromise any name server and only eavesdrops
on messages to learn target-related information in the DNS query. We
are not interested in this type of adversary, since standard security
tools, such as SSL/TLS and/or IPSec \cite{rfc4033}, are effective against
them. Insiders refer to malicious local or authoritative domain name
servers. Although a malicious local name server can be trivially bypassed
by letting clients running their own recursive resolvers, we still
need to seriously analyze the malicious local name server's advantage
given the fact that few clients are willing to sacrifice the performance
gained from caching at the local name server. Insiders can be further
classified into:
\begin{enumerate}
\item Passive (Honest-but-Curious) Insider -  only listens to
  queries sent to it but responds honestly.
\item Active Insider - may drop, forge or manipulate DNS
  responses.
\end{enumerate}
We further assume that malicious name servers under the same namespace
operator's control may collude. 

\subsection{Advantage of the Adversary\label{sec:adv_adv}}
We now formally define the advantage of the adversary $\mathcal{A}$.
First, we give some definitions. ${\eta}_t(d)$ denotes the
average number of queries for name $d$ within time interval $t$. We
give $\eta_t(d)$ different meanings based on where it is measured. At
a local name server, $\eta_t(d)$ represents the number of queries only
from its local administrative domains, while, at an authoritative name
server, $\eta_t(d)$ represents the number of queries from all over the
world.


The advantage of a specific $\mathcal{A}$ in terms of each privacy leak
in Sec.~\ref{sec:privacy_leak} is defined to be:
\begin{enumerate}
\item $\mathbf{Adv}_1(d)$: the probability of $\mathcal{A}$
  successfully outputting the target identity for a query against domain name $d$.
\item $\mathbf{Adv}_2(d, t)$: the probability of $\mathcal{A}$
  successfully outputting $\eta_t(d)$ for a given domain name $d$.
\item $\mathbf{Adv}_3(d_1, d_2, t)$: the probability of $\mathcal{A}$
  successfully outputting $\eta_t(d_1)-\eta_t(d_2)$ for two given
  domain names $d_1$ and $d_2$.
\end{enumerate}
In the following, we may omit the parameter and use $\mathbf{Adv}_1,
\mathbf{Adv}_2, \mathbf{Adv}_3$ for simplicity.






\section{Privacy Threats with Existing Approach \label{sec:related}}
It is easy to see that, in current DNS, all three advantages
in Sec.~\ref{sec:adv_adv} are $1$ for the respective insider adversaries.
In this section, we discuss some existing approaches
that attempt to mitigate the situation.


\subsection{General-Purpose Anonymity Service}
TOR~\cite{tor} uses the onion routing technique to hide traffic source.
As mentioned earlier, it is an effective approach for hiding the source/target relationship,
query source, source query volume and relative source query volume.
However it has nothing to do with mitigating target-related leaks.

\subsection{Random-Set Query \label{sec:random_set}}
Zhao, \textit{et al.}~\cite{Zhao_dns_rangequery} and Castillo-Perez,
\textit{et al.}~\cite{sergio_dnsprivacy} propose a random-set query
approach. Specifically, each time a client queries a domain name $d$,
it constructs a query set $R(d)$, of size $m$, comprising of $d$ ({\it
  real target}) and $m-1$ randomly picked names ({\it confusing
  targets}). The source then queries each of the $m$ names. This
method has the advantage of easy implementation and requires no
changes to current DNS infrastructure. However it has some notable
drawbacks.

\begin{figure}[!t]
\centering
\includegraphics[width=2.5in]{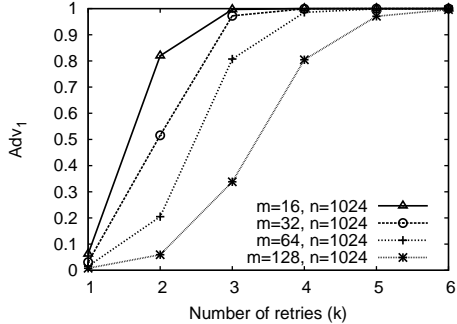}
\caption{Random-set query: $\mathbf{Adv}_1$ increases  dramatically with the number of requests for the same target.} \label{fig:random_set}
\end{figure}

With respect to $\mathbf{Adv}_1(d)$, both \cite{Zhao_dns_rangequery} and
\cite{sergio_dnsprivacy}   claim their respective approaches can achieve
$\frac{1}{m}$. However, this is not necessarily true in the face of an
active malicious local name server that can force a client in its
administrative domain to re-launch a random-set query to the same
target. A local name server can achieve this (while remaining undetected)
by simply dropping requests from a local client within a short period of time. 
Let $R_i(d)$ denote the random set of the $i$th query to
target $d$. If the local DNS server can get $k$ random set queries for
the same target, $R_1(d), \cdots, R_k(d)$, then, since $d \in
\cap_{i=1}^kR_i(d)$, $\mathbf{Adv}_1 = \frac{1}{\mid
  \cap_{i=1}^kR_i(d)\mid}$ instead of $\frac{1}{m}$. Let $S$ denote
the pool of domain names where the client picks both the real target
and the confusing targets. Let $n$ denote the size of $S$. For each
name $d' \neq d  \in S$, the probability of $d'$ appearing
in $R_i(d)$ is $\frac{m-1}{n-1}$. Therefore, the probability of $d' \in \cap_{i=1}^k R_i(d)$ is
$(\frac{m-1}{n-1})^k$ assuming
that confusing targets are independent for each $R_i(d)$.  Thus, the
average size of the intersection $\cap_{i=1}^k R_i(d)$ is
$\frac{(m-1)^k}{(n-1)^{k-1}}+1$.  Fig. \ref{fig:random_set} gives an
example of the relationship between $\mathbf{Adv}_1$ and $k$. It shows that,
as $k$ increases, $\mathbf{Adv}_1$ increases dramatically, especially, when $\frac{m}{n}$ is small.

As far as $\mathbf{Adv}_2(d, t)$ is concerned, $\mathcal{A}$ at the
local name server still has full advantage.  Assume that
$\mathcal{A}$, within the past interval $t$, witnessed $n$ domain
names being queried. $\mathcal{A}$ counts the number of queries from
its administrative domain to $d_i$ in interval $t$ as $q_t(d_i),
\forall i \in [1,n]$ which includes both queries for $d_i$ as real
targets and those querying $d_i$ as confusing targets. Let $\sigma =
\sum_{i=1}^n q_t(d_i)$. Then $\sum_{i=1}^n \eta_t(d_i) =
\frac{\sigma}{m}$ where $\eta_t(d_i)$ is the number of queries which
query $d_i$ as a real target. For each $R(d_j)$, its confusing targets
have probability $\frac{m-1}{n-1}$ of hitting $d_i$ for $i \neq
j$. Therefore, we have:
\begin{align*}
 q_t(d_i) &= \eta_t(d_i)+\frac{m-1}{n-1}\cdot \sum_{j \neq i}\eta_t(d_j)\\
& =  \eta_t(d_i) + \frac{m-1}{n-1}\cdot \left(\frac{\sigma}{m} - \eta_t(d_i))\right)
\end{align*}
Solving this equation yields $\eta_t(d_i) = \frac{q_t(d_i)(n-1)}{(n-m)} - \frac{\sigma(m-1)}{m(n-m)}$.

We also observe that $\mathbf{Adv}_3(d_a, d_b, t)=1$ for both malicious
local name servers and authoritative name servers. It is trivial for
$\mathcal{A}$ at a local name server to compute
$\eta_t(d_a)-\eta_t(d_b)$, since it already knows $\eta_t(d_a)$ and
$\eta_t(d_b)$ ($\mathbf{Adv}_2 = 1$).  If $\mathcal{A}$ is an
authoritative server, it does not learn $\eta_t(d_i)$, since it does not
know $\sigma$ and $n$. However,
$\eta_t(d_a) - \eta_t(d_b) = \frac{(n-1)(q_t(d_a)-q_t(d_b))}{(n-m)} \approx
{q_t(d_a)-q_t(d_b)}$, given $n \gg m$.  Thus, $\mathcal{A}$ can also easily
determine $\eta_t(d_a) - \eta_t(d_b)$.

\subsection{Fixed-Set Query}
Instead of choosing $m-1$ confusing targets randomly, each client can
fix them for each real target $d$. This can be achieved efficiently
through a pseudo-random function with a seed only known by the
client. This way, each query for the same target will end up with the
same set and $\mathbf{Adv}_1 = \frac{1}{m}$. However, this approach
lacks robustness. If the adversary learns the mapping of a target $d$
from one client, all future requests for $d$ from that client is clear
to the adversary since the mapping is fixed. Moreover, for a malicious
authoritative name server, $\mathbf{Adv}_3(d_a, d_b, t)$ is still
$1$. This is because it is hard to synchronize the mapping function
between different clients and therefore the fixed set chosen for the
same target by different clients are still independent. If in time
$t$, the number of requests to $d_a$ or $d_b$ from the same client is
negligible to the total volume, then $q_t(d_a)-q_t(d_b)$ still gives a
good estimate of $\eta_t(d_a)-\eta_t(d_b)$ for the same reason as
discussed in Sec.~\ref{sec:random_set}.

\subsection{Combining Tor and Random Set Query}

\begin{figure}[!t]
\centering
\includegraphics[width=5in]{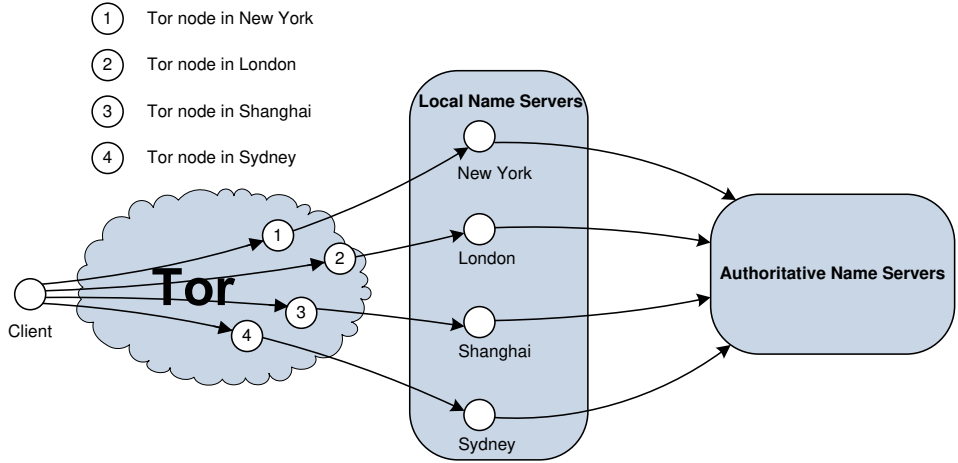}
\caption{Client poses Random-Set Query to local name servers at different locations} \label{fig:tor}
\end{figure}

If we combine Random-Set Query with TOR, the adversary at the local
name server cannot force the same client to relaunch a random-set
query to the same target.  As shown in Fig. \ref{fig:tor}, a client
can use TOR to query ``local'' name servers around the world by
forwarding random-set queries to TOR nodes in different regions. This
way, an adversary at a single local name server cannot see multiple
random-set queries to the same target by dropping client's requests,
since if a random-set query is dropped, the client can choose a
different local name server to relaunch the query.  As long as there
are honest local name servers, the client's query can be answered.
This approach effectively reduces $\mathbf{Adv}_1$ to $\frac{1}{m}$
for adversaries at local name servers. However, this gain does not
come for free. To maintain perfect unlinkability, the client needs to
create a new TOR circuit, which can take more than $4$ seconds
\cite{Panchenko08tor}, for each DNS query. This is a very high
overhead for a client that only wants to hide its query target. Also,
if the number of TOR users in a local domain is negligible,
$\mathbf{Adv}_2$ and $\mathbf{Adv}_3$ for adversaries at local name
servers remain quite high. Moreover, even there are large number of
TOR users, $\mathbf{Adv}_3$ for adversaries at authoritative servers
is not affected, since they are the ultimate ``sink'' for all queries.



\subsection{DNS Caching}
DNS caching only moves the overhead of DNS
resolution from authoritative name servers to local name
servers. Thus, as pertains to  $\mathcal{A}$ at a local name server,
caching does not offer any privacy benefits.

With caching, if $\mathcal{A}$ is located at an authoritative name
server, it cannot learn the exact query volume for a target. However,
it can still compute the average target query volume. To see this, we
first need to model the arrival process of DNS request. Poisson
processes are widely used to model events if times between consecutive
events are independent random variables and the number of events in
one interval is independent from previous intervals. Earlier
work~\cite{Paxson95wide-areatraffic} suggests that session
inter-arrivals can be reasonably approximated by an exponential
distribution, and that has been supported in studies on DNS
caching~\cite{jung2003modeling}. We thus consider it
reasonable to model DNS requests' arrival events as a poisson process.

We assume the arrival process of DNS queries to target $d$ from a local
administrative domain $l$ conforms to a poisson distribution with
arrival rate $\lambda_l(d)$. Let $ttl(d)$ denote the TTL for  $d$.
After a client within a local administrative domain
queries its local name server for $d$, the local name server
caches the response (from authoritative name server) for time
$ttl(d)$. After seeing a query against $d$
from a local name server from domain $l$, the adversary knows that, within
the next period of length $ttl(d)$, it will not see further queries from this local name
server. However, because of poisson distribution, $\mathcal{A}$ knows the
average number of queries against $d$ in the following $ttl(d)$ time
is $\lambda_l(d) \cdot ttl(d)$. Given a longer time period $t > ttl(d)$, $\mathcal{A}$
can count the number of queries, $k_l(d)$, against $d$ from
the local name server. Then, it can estimate the total number of
queries that occur in time $t$ as:  $\lambda_l(d) \cdot ttl(d) \cdot
k_l(d)+ k_l(d)$. 
Again, due to poisson distribution, the following equation holds:
\begin{align*}
\lambda_l(d) \cdot t = \lambda_l(d) \cdot ttl(d) \cdot k_l(d) + k_l(d)
\end{align*}
Solving it yields $\lambda_l(d) = \frac{k_l(d)}{t-ttl(d)\cdot
  k_l(d)}$ and $t\cdot \lambda_l(d) = \frac{tk_l(d)}{t-ttl(d)\cdot
  k_l(d)}$ as the average number of queries from that local
administrative domain $l$. Finally, summing up the average number
of queries from all local domains, $\mathcal{A}$ obtains
$\eta_t(d)=\sum_{l}\frac{tk_l(d)}{t-ttl(d)\cdot k_l(d)}$.

If $\mathcal{A}$ is an active insider at an authoritative name
server, it can also set the TTL to a negligible value,
since it has the right to do so. Then, $\mathcal{A}$ retains the
same advantage as discussed in previous sections.

\subsection{Information-theoretic Private Information Retrieval }
Information-theoretic PIR is a cryptographic technique that allows a client
to retrieve a record from two or more non-colluding servers, each having
a copy of the database, without either server learning which record the
client wants to retrieve. Zhao, et al. \cite{zhao_dns_pir} use the
two-server version of information-theoretic PIR to achieve DNS query
privacy. However, information-theoretic PIR requires the assumption of
multiple non-colluding servers, each having a copy of the database. In
reality, servers with duplicate database are usually the primary and
secondary servers for the same zone. These servers can collude with
each other according to the assumptions in Sec.~\ref{sec:adv_type},
since they are under control of the same namespace operators. Thus,
we do not consider
information-theoretic PIR to be suitable for preventing leaks in DNS
queries.



\section{Building Blocks \label{sec:blocks} }
In this section, we introduce some building blocks that are used in our system.

\subsection{DHT-based DNS \label{sec:codons}}
%
%
Distributed hash tables (DHTs) are a class of decentralized
distributed systems that provide a lookup service similar to a hash
table: (key, value) pairs are stored in the DHT, and any participating
node can efficiently retrieve the value associated with a given
key. There has been plenty of research devoted to flat-structure
DHT-based DNS \cite{CoDoNS, Cox2002Serving}. In fact, some results
have already seen trial deployment on the Internet. DHT-based DNS
solutions propose to use a flat-structure to replace the current
hierarchical DNS structure. Basically, it is a peer-to-peer domain
name system wherein both nodes and objects have randomly assigned
identifiers from the same circular space. Here we take
CoDoNS~\cite{CoDoNS} as an example and describe its architecture in
detail.

CoDoNS is based on prefix-matching DHT. It designates the node whose
identifier is closest to the consistent hash of a domain name as the
\textit{home node} for that domain name. The home node stores a
permanent copy of the resource records owned by that domain name and
manages their replication. If the home node fails, the next closest
node in the identifier space automatically becomes the new home node.


CoDoNS supports DNSSEC by separating the authentication of data from
the service of that data. Every namespace operator has a
public-private key pair; the private key is used to digitally sign
DNS records managed by that operator, and the corresponding public
key is in turn certified by a signature from a domain higher up in
the hierarchy. This process creates a chain of certificates,
terminating at a small number of well-known public keys for globally
trusted authorities. The signature and the public key are stored in
CoDoNS as resource records of type sig and key respectively. Clients
can verify the authenticity of a resource record by fetching the sig
record and the key record from the CoDoNS. CoDoNS servers cache the
certificates along with the resource records to help clients check
the validity of a record.

CoDoNS decouples namespace management from the physical location of
name servers in the network. Namespace operators do not need to
participate in the data serving. Their responsibility is only to sell
certificates of names they account for to nameowners. Nameowners later
introduce these names into CoDoNS. CoDoNS servers authenticate
nameowners directly through certificates provided for every insertion,
deletion and update. CoDoNS is agnostic about the hierarchical
structure of namespace while namespace operators are agnostic about
which CoDoNS servers are serving the names they give a certificate
to.

Some common DNS operations are not specified by CoDoNS but can be
easily combined into any DHT-based DNS. For example, to support
wildcard query, a DHT node, when finding unmatched result for a
specific subdomain name, e.g. ``random.example.com'', the DHT node can
redirect this request to a node accounting for "*.example.com" where
there is a record for default mapping.

\subsection{Gentry-Ramzan(GR)  cPIR \label{gentry_pir}}
Gentry-Ramzan cPIR~\cite{Gentry2005SingleDatabase} is a
computational private information retrieval (cPIR) scheme. cPIR
schemes typically assume a single server, in contrast to  multiple
non-colluding replicated servers  required by information-theoretic
PIR.

In Gentry-Ramzan  (GR) cPIR, the database server and clients share $t$
distinct prime numbers $\{p_1, \dots, p_t\}$, and $t$ prime powers
$\{\pi_1, \dots, \pi_t\}$ where $\pi_i=p_i^{c_i}$ and $c_i=\lceil
l/\log_2 p_i \rceil$ for a database with $t$ blocks, each of size at
most $l$.  Each block $C_i$ is associated with a prime power
$\pi_i$. Using the Chinese Remainder Theorem, the server can express
the entire database as an integer $e$ that satisfies $e \equiv
C_i(\text{mod}\; \pi_i)$. Notice that to retrieve $C_i$, it suffices
to retrieve $e \mod \pi_i$ since $0 \le C_i < 2^l \le \pi_i$.

\textbf{Query Generation:} The user determines index $i$ of its
desired block, and then generates an appropriate cyclic group $G =
\langle g \rangle$ with order $|G| = q \pi_i$ for some suitable
integer $q$. It sends $(G, g)$ to the server and keeps $q$ private. It
also stores $h=g^q$ for future use. Note that $h$ is a generator of
subgroup $H$ of order $\pi_i$.

\textbf{Database Response Generation:} The database server first
expresses each block $C_j$ as a number in $[0, 2^l-1]$ in the obvious
fashion. Then it sets $e$ to be the smallest positive integer such that
$e \equiv C_i(\text{mod}\; \pi_i)$ for all $i$ by Chinese Remainder
Theorem. Last it outputs the response $g_e = g^e \in G$ to the user.

\textbf{Response Retrieval:} Given the input $(\pi, g_e, G, q, h)$,
the user retrieves block $C_i$ as follows: It computes $h_e = g_e^q$
and gets $C_i$ through the discrete logarithm $\log_h h_e$ within the
subgroup $H \subset G$ of order $\pi_i = p_i^{c_i}$ using
Pohlig-Hellman algorithm.

To instantiate the cyclic group $G$, one may first construct a random
``semi-safe'' prime $Q_0 = 2q_0 \pi_i + 1$ for some prime $q_0$ and a
random safe prime $Q_1 = 2q_1 + 1$ for some prime $q_1$ and set $m =
Q_0Q_1$ as a composite modulus. Then one can take a quadratic residue
from $\mathbf{Z}_m^*$ and test if it is a generator for a cyclic group
whose order is $q_0q_1\pi_i$ which is $\phi(m)/4$ where $\phi$ is
Euler's totient function until a generator is found.

The security of this scheme comes from $\Phi$-Hiding assumption that
it is computationally intractable to decide whether a given small
prime divides $\phi(m)$ where $m$ is a composite integer of unknown
factorization. In other words, given a cyclic group $G$ constructed in
above way where $|G|=\phi(m)/4$, the server is computationally
difficult to test which $\pi_i$ divides $|G|$ and therefore
computationally difficult to know which element the user tries to retrieve.





\section{PPDNS Architecture\label{sec:architecture}}
This section describes the Privacy Preserving Domain Name
System (PPDNS) which aims to mitigate target-related DNS
privacy problems. PPDNS is built upon a DHT-based naming infrastructure. It takes
advantage of DHT's index to implement
fixed-range routing. It also offers bandwidth-limited clients
the option of using cPIR to reduce communication overhead.

\subsection{Why DHT-based DNS?}
Having examined existing approaches, we conclude that current DNS
cannot be amended to offer privacy due to its fundamental design
features. Therefore, we must consider new approaches for emulating DNS
that would lend themselves to better privacy. We believe that
DHT-based DNS is a viable candidate for a privacy-preserving domain
name system.

There are two privacy advantages of DHT-based DNS. First, it
uses consistent hashing for object
location, which enables approaches, such as fixed range query and
cPIR. Second, the flat and balanced storage structure
gives equal power to adversaries compromising different servers. This
is in contrast to adversaries in current DNS that attain more power by
compromising higher-level DNS servers or servers with more
records.

\subsection{Why GR cPIR? \label{sec:why_gentrypir}}
There are many available cPIR schemes. Our choice of GR
cPIR is based on two factors. First, its communication complexity is
$\mathcal{O}(k+d)$, where $k \le \log n$ is a security parameter that
depends on the database size $n$ and $d$ is the bit-length of the
retrieved database block. This communication complexity is, by far, the
best. Second, the dominant component of server's
computation is in computing $g^e$ which consists of at most
$2(t\cdot l)$ modular multiplications, where $t$ is the number of
blocks in server's database and $l$ is the block length. As shown
in Sec. \ref{sec:gentry_perform} below, this amount of overhead
can be tolerable for DNS servers in some circumstances.

\subsection{PPDNS protocol \label{sec:ppdns}}
%
%
We begin by introducing some definitions and notation. All clients and
PPDNS nodes share a hash function that maps any domain name into a
circular space $[0, N-1]$ where $N$ is the upper-bound of the hash
space. We use SHA1~\cite{2002-sha} as the hash function, and thus,
$N = 2^{160}$. We say that an identifier is {\it empty} if
there are no domain names attached to it. We refer to the identifier
space density as $\rho$ -- the ratio of the number of non-empty
identifiers to the total number of identifiers. Each PPDNS node can
compute $\rho$ individually, due to the uniformity of the
hash function.  We define query range for a {\it target identifier} to
be a set of continuous identifiers, including the target identifier. We
refer to the identifiers in the query range -- except the target --
as {\it confusing identifiers}. Each client has a security
parameter $m$ representing the number of non-empty identifiers it expects
in the query range.


Each client gets its estimate of $\rho$ from its local name server and
periodically updates it as a moving average of its previous
estimate and newly learnt $\rho$ values. A client who wants to
resolve a target domain name first hashes it into an identifier
$i^*$. Then, it determines the number of non-empty identifiers,
$m$, in its query range. The actual range size is computed as
$2^s$ where $s = \lceil \log_2 \frac{m}{\rho} \rceil$. We expect most
clients to have the same $m$ and $\rho$. However, we allow clients
to have different $m$-s and have some estimation error on
$\rho$. The reason for the client to select its range as a power
of $2$ is to accommodate these differences. Finally the query range
for target identifier $i^*$ is formulated as:
\begin{align}
Q(i^*) = \left[\left\lfloor
  \frac{i^*}{2^s} \right\rfloor \cdot 2^s,\;\; \left(\left\lfloor \frac{i^*}{2^s}
  \right\rfloor+1\right) \cdot 2^s-1 \right] \label{eq:range_rule}
\end{align}
Theorem~\ref{thm:th1} shows that, if the query range is formed in the
above manner, then the query range generated for each
identifier is fixed and the same query range can be posed for each
identifier inside this range. Theorem~\ref{thm:th2} shows that, even for
clients with different $m$ and $\rho$ values, the intersection of their
query ranges is guaranteed by the smallest query range among them.

Next, the client constructs a range query, which only needs to include
the start and end identifiers. The range query is initially sent to
the local PPDNS node which replies immediately if it has a cached copy
of the queried range. Otherwise, it routes the query towards its
destination node, which accounts for the identifiers inside the range,
by following the underlying DHT protocol.  If any intermediate PPDNS
nodes happen to have a cache of the queried range, they respond
directly to the local PPDNS node and the query stops. Otherwise, they
pass the query onto the next intermediate node until it reaches the
destination.

Range splitting may occur and multiple sub-range queries may be
generated at intermediate PPDNS nodes. A PPDNS node that has complete
records for a sub-range query returns all records (with identifiers
inside the subrange) to the local PPDNS server. The local PPDNS
server waits until all responses
arrive and then delegates them to the client. The local
PPDNS server also caches the whole range of responses for the minimum
TTL among all responses in the range. Since all clients in the
same administrative domain get $\rho$ from the same local name
server, they come up with the same range size, as long as $m$ is also the
same. Therefore, by caching the whole range answered to one client, the local
PPDNS server can also answer later queries from other clients.

\begin{figure}[!t]
\centering
\includegraphics[width=3in]{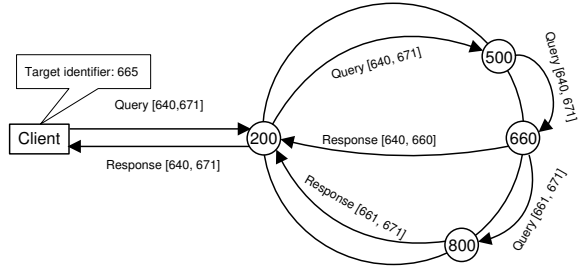}
\caption{Range Query in DHT-based DNS} \label{fig:range_split}
\end{figure}

As an example, consider Chord~\cite{Stoica2001Chord} circular space
shown in Fig. \ref{fig:range_split} where $\rho=\frac{1}{2}, m=16,
N=1024$. Suppose the client wants to query a domain name with hash
identifier $665$. Then the client will make a range query of $[640,
671]$ according to Eq. (\ref{eq:range_rule}) to its local PPDNS
server (node $200$). Node $200$ forwards this range query to node
$500$ by looking at its finger table.  Node $500$ further forwards
the range query to node $660$ which is the home node for range
$[640, 660]$. Node $660$ first splits the range into $[640, 660]$ and
$[661, 671]$. Then it replies the records corresponding to range
$[640, 660]$ to the local PPDNS server and further forwards the
subrange query of $[661, 671]$ to node $800$ which later replies to
the local name server with records corresponding to range $[661,
671]$. The local name server finally delegates the answer to the
client.
\subsection{cPIR support}
\begin{figure*}[!t]
\centering
\includegraphics[width=5in]{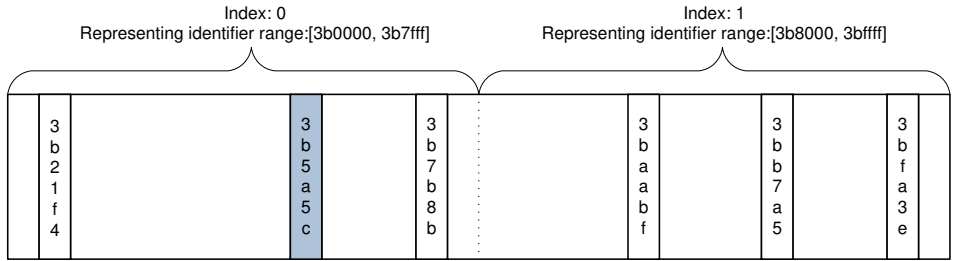}
\caption{Block index decision where $n_{pir} = 2$. The number inside
  the block shows hash identifier. The blue one is the one the client
  wants to retrieve.}
\label{fig:pir_block}
\end{figure*}

As an option, the client may choose to use cPIR to save bandwidth.
Modern residential ISPs can offer up to $6$Mbps
download speeds; thus, most clients do not need to worry about the extra
communication overhead due to range queries. However, for
GPRS~\cite{gprs} users in developing countries, the maximum currently
achievable  speed is between $56$ and $114$kbps, depending on the
distance to the base station.
If we assume the user issues one range query with $m=128$ every $4$
seconds,  the bandwidth tallies up to $16$kbps
for regular DNS ($512$ bytes per response) and $64$kbps for DNSSEC
($2048$ bytes per response), which consumes a lot of the channel's
capacity. This is unacceptable for most GPRS users.
Moreover, many cell phone users' data plans are based on traffic volume,
especially, when roaming in foreign countries. These factors motivate
the use of cPIR.

To take advantage of cPIR, the client needs to view the query range as
a small database. An output of a secure hash function is typically on the order of
hundreds of bits (e.g. 160 for SHA-1). The total number of
domain names with respect to the total number of identifiers in the
hash space is quite low,  i.e., $\rho$ is very small.
Given the low density of the hash space, simply using the
hash identifier as the index into the database for cPIR would work.
However, it
would cause very high computation and bandwidth overhead as cPIR's
overhead is proportional to the number of database entries.

One way to tackle this problem is for the client to
divide the range (database) into $n_{pir}$ equal blocks and assign
a consecutive index to each. Then, the client uses cPIR to retrieve
the block which contains the desired identifier's record.  Because the
hash function is near-uniform, the average number of
non-empty records in each block is $m/n_{pir}$. This way, cPIR
overhead is proportional to $m$ instead of
$\frac{m}{\rho}$.

%
Consider an example in Fig. \ref{fig:pir_block} where hash output is
$24$ bits and $m=6$. A client picks a query range  (3b0000-3bffff).
It wants to retrieve
the blue record with identifier 3b5a5c but it does not know the
other non-empty identifiers. If the client sets $n_{pir}=2$,
the whole range is split into two equal-size blocks and, by reassigning
the blocks, it knows that
the blue record
lies in the first reassigned block which accounts for identifiers from
$0$ to 3b7fff. Next, the client retrieves the first reassigned block, which
includes three original records from the server. The choice of
$n_{pir}$ depends on the computation power of the server, which is to
be analyzed in Sec. \ref{sec:gentry_perform}. The larger $n_{pir}$
is, the lower the communication overhead is while the higher the
computation overhead is. Any cPIR schemes can fit here. Due to the
reasons we explained in Sec.~\ref{sec:why_gentrypir}, we currently
employ GR cPIR in PPDNS. In the future, however, if there
are more efficient cPIR schemes, we can easily apply them here.

In order to incorporate GR cPIR, we need the local
name server and the clients to share a sequence of consequtive prime
numbers. The number of prime numbers should be equal to the maximum
number of blocks the local name server allows a client to retrieve
each time. When the client issues a range query, it also indicates a
range of prime numbers. The client can do so by only showing the
ending prime numbers' index in the consecutive prime numbers' list if
the client always use the first prime it shared with the server as the
starting prime. From the range of primes embedded inside the range
query, the server can compute the number of blocks the client wants to
divide the range into. Then, the local PPDNS server gets all the
responses to this range from other servers (through normal range
query described in Sec. \ref{sec:ppdns}) and then performs cPIR with the client.
If the local PPDNS server happens to have cached the whole range,
it can perform cPIR immediately.

Currently, we only employ cPIR between the client and the local PPDNS
server. 
If cPIR was also used between PPDNS servers, the range cache
would be no longer possible. We observe through simulation in
Sec.~\ref{sec:simulation} that the cache captures more than $50$\% of all
range queries. Thus, giving up the cache actually counteracts cPIR benefits.
Moreover, due to the high capacity of backbone links,
backbone traffic resulting from a range query is very limited (less than
$1$\% in our simulations). Therefore, we see no reason to use cPIR
between PPDNS servers.

As an example in Fig.~\ref{fig:range_split}, if the client chooses to
use GR cPIR and decides to divide the range into $8$ blocks
($n_{pir}=8$), it indicates its prime range by sending the prime index
$8$ along with its range query $[640, 671]$, which means using
the first eight prime numbers in the consecutive prime number lists
shared with the local name server to execute GR cPIR. When
the client receives the answer to the query range $[640, 671]$, it
knows its desired index $665$ can be extracted by using the $7$th
prime number in the shared prime number lists.

\subsection{Other Ideas \label{sec:otherthoughts}}
To facilitate incremental deployment, the client relies on current DNS
to resolve queries for records not in PPDNS and explicitly inserts
them into the system. To protect its privacy from adversaries in both
DNS and PPDNS, the client first queries PPDNS with the range formed by
Eq.~\ref{eq:range_rule}.  Once client realizes the target domain
name's RR set is not inside the response, it uses the same range query
to query current DNS. Instead, the domain names except the real target
inside the range query to current DNS is learnt from the responses of
PPDNS.  Once the client learns the records for target domain name, it
asks a random PPDNS node (e.g., chosen by random
walk~\cite{Gkantsidis2004Random}) to insert the target into the
system. DNSSEC is employed for preventing poisoning attack.

\section{Security Analysis \label{sec:analysis}}
In this section, we analyze the security of PPDNS in terms of adversary's
advantage described in Section~\ref{sec:adv_adv}.

First, we state two theorems about range selection in
Eq.~\ref{eq:range_rule}.
In short, the first theorem states that the query
range for a specific identifier is fixed, such that the same
range can be formed for each identifier inside this range. The second
states that even if clients have different $m$ and
$\rho$ values, the size of the intersection of their ranges for the same
identifier cannot be smaller than the smallest range size among them,
which gives the adversary limited power by making intersection over query range.
\begin{theorem}\label{thm:th1}
  $\forall s \in \mathbf{Z}$ such that $2^s$ divides N, the query
  range generated by Eq.~(\ref{eq:range_rule}) for any identifier is
  fixed and the same query range can be posed for any identifier
  inside this range.

\end{theorem}
\begin{proof}
  Suppose $N = k\cdot 2^s$. Then, for any identifier $i \in [0, N-1]$,
  there exists one $j \in [0, k-1]$ such that $i \in [j\cdot 2^s,\;
  (j+1)\cdot 2^s-1]$. Moreover, $\forall i \in [j\cdot 2^s,\;
  (j+1)\cdot 2^s-1]$, the range generated by Eq.~(\ref{eq:range_rule})
  for $i$ is $[j\cdot 2^s,\; (j+1)\cdot 2^s-1]$.
\end{proof}
\begin{theorem} \label{thm:th2}
  The lower bound of the size of the intersection for all query
  ranges generated for a common identifier by different clients
  with different $m$ and $\rho$ is guaranteed by the smallest $ 2^{
    \lceil \log_2 \frac{m}{\rho} \rceil}$.
\end{theorem}
\begin{proof}
  We only need to prove that, if two ranges:
  $\gamma_1 = [k_1\cdot 2^{s_1}, (k_1+1)\cdot 2^{s_1}-1]$ and
  $\gamma_2 = [k_2\cdot 2^{s_2}, (k_2+1)\cdot 2^{s_2}-1]$ where
  $s_2>s_1, k_1, k_2, s_1, s_2 \in \mathbf{Z}^+$, overlap, then
  $\gamma_1 \subseteq \gamma_2$.

  If $k_2\cdot 2^{s_2} \le k_1\cdot 2^{s_1} < (k_2+1)\cdot 2^{s_2}-1$,
  then $k_1 < (k_2+1)2^{s_2-s_1}$. Since both sides are integers, we
  have $k_1 \le (k_2+1)2^{s_2-s_1} - 1$. Therefore, $k_2\cdot 2^{s_2} <
  (k_1+1)\cdot 2^{s_1}-1 \le (k_2+1)\cdot 2^{s_2}-1$. Similarly, we can
  prove that if $k_2\cdot 2^{s_2} < (k_1+1)\cdot 2^{s_1}-1 \le
  (k_2+1)\cdot 2^{s_2}-1$, then $k_2\cdot 2^{s_2} \le k_1\cdot 2^{s_1}
  < (k_2+1)\cdot 2^{s_2}-1$.

  Thus, if one end of $\gamma_1$ falls inside $\gamma_2$, so does
  the other. Then, the theorem holds.
\end{proof}

We expect that most clients have the same security parameter $m$. As
to be shown in Sec.~\ref{sec:density_perform}, the density estimated
at different PPDNS node in real senario is very close. These facts
combined with Theorem~\ref{thm:th2} means the intersection size
should be reasonably close to $\frac{m}{\rho}$.  Even if some
clients use extremely small $m$, it only affects $\mathbf{Adv}_1$ as
to its own query and has limited effect on $\mathbf{Adv}_2$ and
$\mathbf{Adv}_3$ as long as the number of such clients are small.



Now we use the model of Sec.~\ref{sec:adv_model} to analyze the
security of PPDNS. We assume all clients have the same $m$ and $\rho$ parameters.
Recall that any
PPDNS server can be both a local name server and an authoritative name
server. We neglect the security analysis for intermediate routing
PPDNS nodes, since their advantage is captured by the authoritative
name server.


A passive $\mathcal{A}$ at a local PPDNS server receives range
queries from, and provides responses, to clients in its
administrative domain. By examining range responses that it
delegates to clients, it knows all domain names inside the range.
Therefore, $\mathbf{Adv}_1 = \frac{1}{m}$.

As far as target volume, suppose that $V$ is the total volume for
target $d$'s query range $Q(d)$ from the local administrative
domain. Then, the average volume to $d$ should be $p\cdot V$, where
$p$ is the probability of $d$ being queried in each range query to
$Q(d)$. If we assume $\mathcal{A}$ does not have any prior knowledge
of $p$, any value from $0$ to $1$ is equally likely, from
$\mathcal{A}$'s perspective. Therefore, the target volume can be any
value between $0$ and $V$ and $\mathbf{Adv}_2 = \frac{1}{V+1}$. 

One issue may come into mind is that if the confusing identifiers have
low probability and the real one has high probability, a plaintext
attack is possible since hash value is fixed for each
identifier. However, we think this kind of difference between
confusing and the real identifier comes from prior knowledge before
PPDNS is deployed. Once PPDNS is fully deployed, the prior knowledge
would become less and less credible over time.

In terms of
relative target volume, given two domain names $d_a$ and $d_b$, if the
total volume to $Q(d_a)$ and $Q(d_b)$ is $V_a$ and $V_b$ respectively,
then, in $\mathcal{A}$'s view, $\eta_t(d_a)-\eta_t(d_b)$ can be any
value between $-V_b$ and $V_a$. Therefore, $\mathbf{Adv}_3 =
\frac{1}{V_a+V_b + 1}$.  Note that $\mathbf{Adv}_2$ and $\mathbf{Adv}_3$
only relies on the volume and has nothing to do with query range
$m$. In particular, if $V=0$ (resp. $V_a=V_b=0$), $\mathbf{Adv}_2 = 1$
(resp. $\mathbf{Adv}_3 =1$).

An active $\mathcal{A}$ at a local PPDNS server can drop
a response without being detected. However, this does not give it
more advantage, since the same
range query is always issued for the same target, as shown in
Theorem~\ref{thm:th1}. $\mathcal{A}$ has no incentive to
spread false density information to clients in its
administrative domain, as doing so can be detected by
checking the density difference with other PPDNS nodes. $\mathcal{A}$
also has no motivation to forge or manipulate responses since DNSSEC can
effectively prevent this. Therefore, the advantage for
the active malicious local DNS server is the same as that of its
passive counterpart.

A passive $\mathcal{A}$ at an authoritative PPDNS server
knows whether an identifier in its charge is empty or not, since
$\mathbf{Adv}_1 = \frac{1}{m}$. The advantage of guessing target
query volume is again $\mathbf{Adv}_2 = \frac{1}{V+1}$, where $V$ is the
total volume to the target identifier's query range. And, the
advantage of guessing relative target query volume is
$\mathbf{Adv}_3 = \frac{1}{V_1+V_2 + 1}$ if $V_1$, where $V_2$ is the
volume to two given domain names' query range.

An active $\mathcal{A}$ at an authoritative PPDNS server
cannot change the TTL or manipulate RRset, unlike an  authoritative
name server in today's DNS,  since all records are signed by different
namespace operators and the authoritative PPDNS server is just a host for
domain names for which it does not have corresponding signing keys.  Thus,
its advantage is the same as that of a passive adversary.

\section{Performance Evaluation \label{sec:performance_eval}}
This section evaluates the performance of different components comprising PPDNS.

We start with judging the accuracy of computing density $\rho$, which
plays an important role in determining query range.  Then we benchmark
a cPIR scheme, an optional feature supported by PPDNS, to see how much
bandwidth it can save a client and how much computation power it would
take. Last, we test PPDNS's performance under realistic backbone
environment through simulation.


\subsection{Density Estimation Accuracy \label{sec:density_perform}}
Density $\rho$ estimated at each PPDNS node plays an important role in
determining the query range Eq.~\ref{eq:range_rule}. Recall that
clients in the same administrative domain have the same estimated
$\rho$, since they all get it from the local PPDNS server. However,
there is a security threat if $\rho$ values estimated at different
PPDNS nodes vary significantly. This is because: (1) large deviation
in $\rho$ allows the local adversary to cheat clients by spreading
larger $\rho$, which may not be detected
and (2) by intersecting query ranges, $\mathcal{A}$ may gain higher
advantage, since the intersection size is the smallest query range for
a specific target identifier.

SHA1 is well known for maintaining uniformity across its
space. However, due to the importance of $\rho$, we want to make sure
SHA1 can maintain high uniformity in terms of domain names as input.
To test this, we retrieved all the second-level domain names ending
with ``.com'' ($80,044,181$ total) from VeriSign~\cite{verisign} in
July 2009, prefixed each with ``www.'', hashed with SHA1 and stored
these results at Chord Nodes.

\begin{figure*}[!t]
\begin{minipage}[b]{0.33\linewidth} 
\centering
\includegraphics[width=5.3cm]{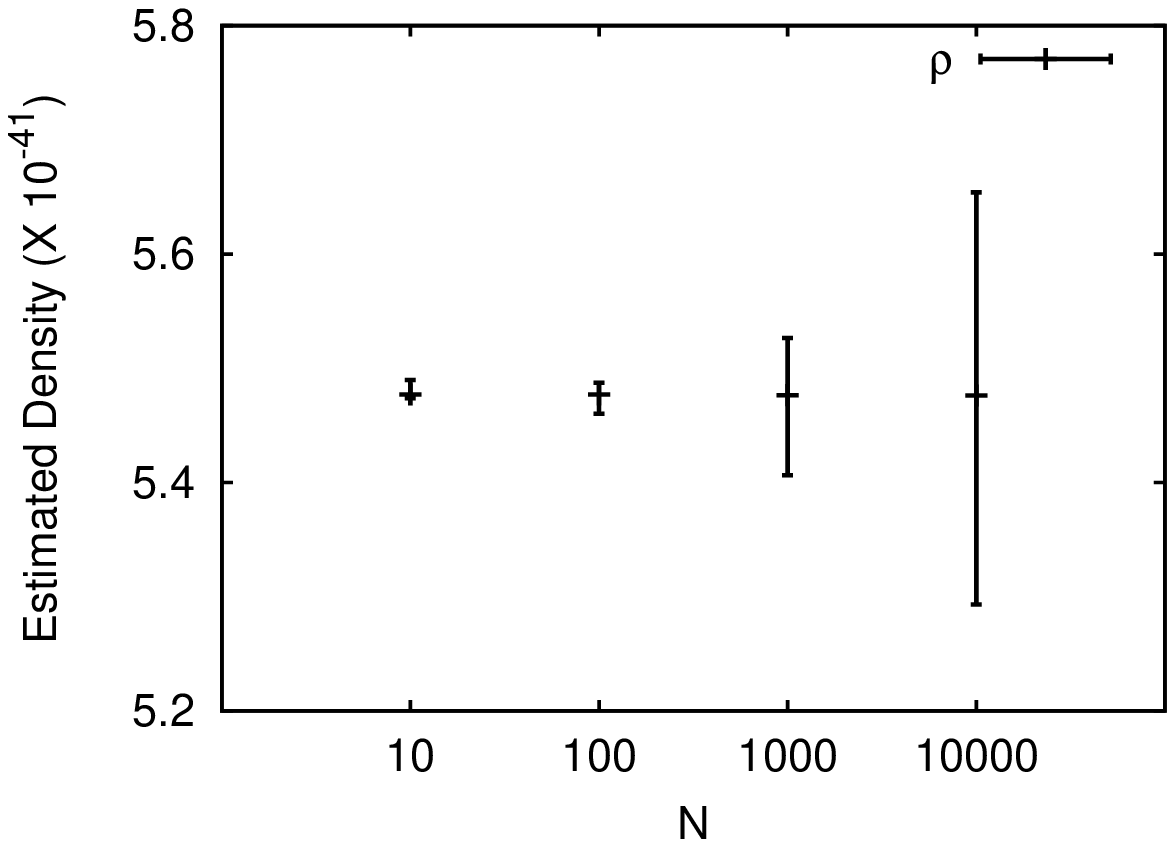}
\caption{Density deviation}
\label{fig:density_dev}
\end{minipage}
\hspace{0.1cm} 
\begin{minipage}[b]{0.33\linewidth}
\centering
\includegraphics[width=5.3cm]{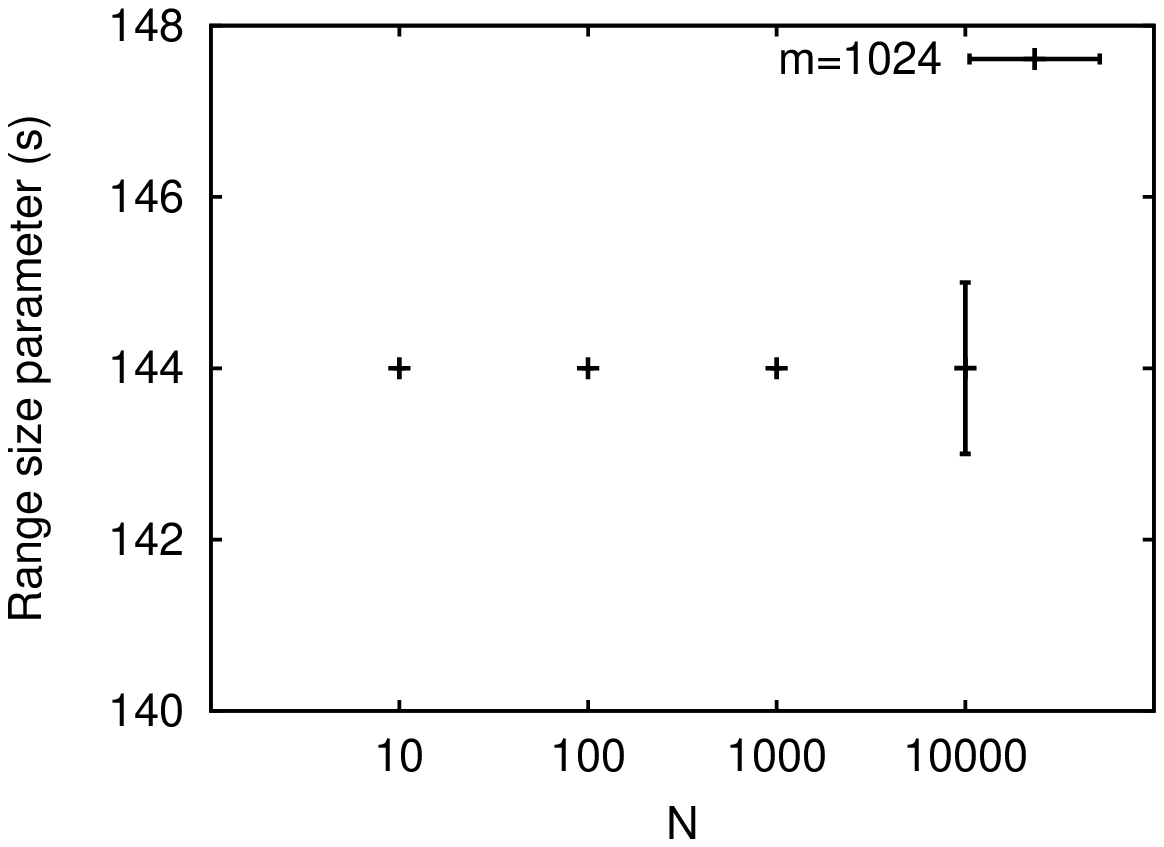}
\caption{Range deviation}
\label{fig:range_dev}
\end{minipage}
\hspace{0.1cm} 
\begin{minipage}[b]{0.33\linewidth}
\centering
\includegraphics[width=5.3cm]{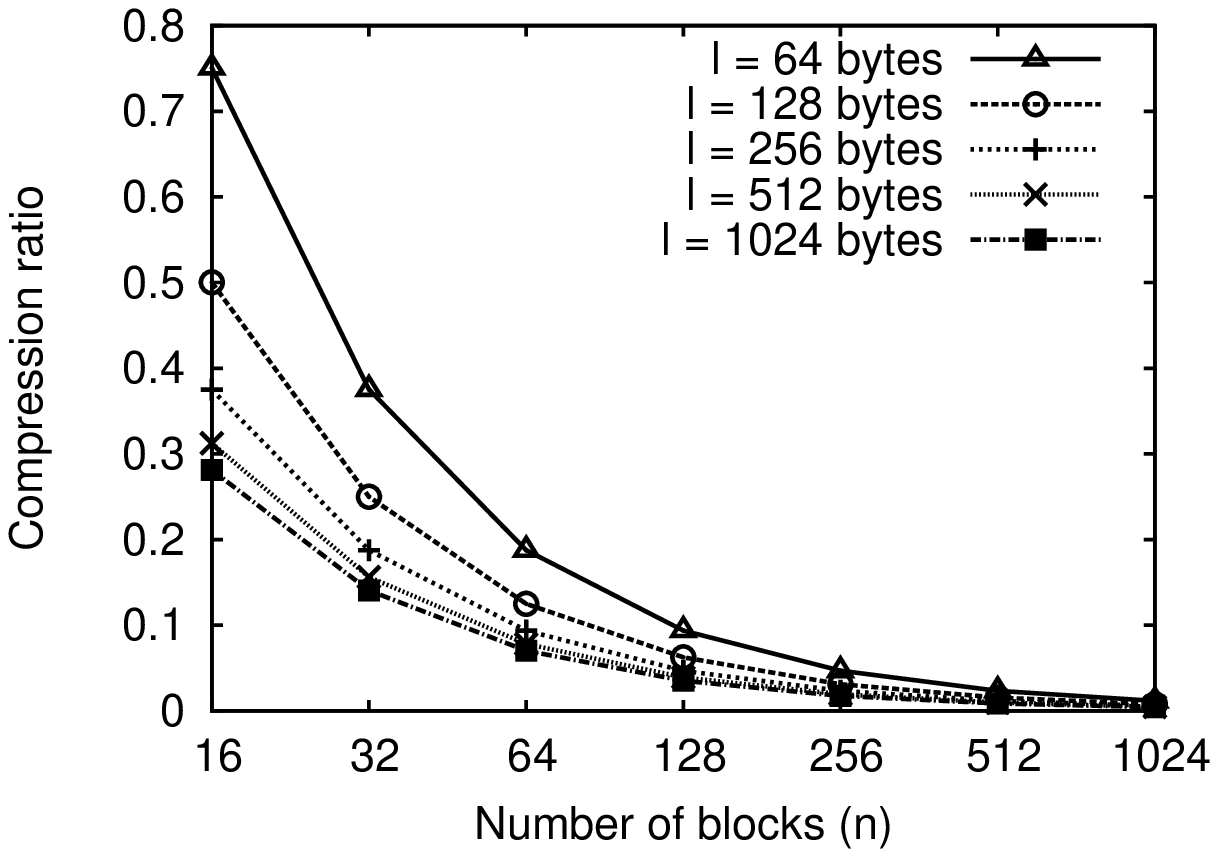}
\caption{Compression Ratio}
\label{fig:gentry_compression}
\end{minipage}
\end{figure*}

Fig. \ref{fig:density_dev} shows the density deviation with respect
to the number of PPDNS. Errorbars show quantiles
5, 50 and 95 related to the set of density estimation made at each
node. As we can see, the deviation of density estimation increases
with the number of nodes. This is expected because, with more nodes,
the sample space at each node becomes smaller.

To measure the range deviation, we set $m$ to be $1024$ for all nodes
and check the deviation of $s=\lceil \log_2 \frac{m}{\rho} \rceil$
which is the logarithm of the final range size to the base of two. We
choose a large $m$ here because larger $m$ tends to generate larger
error. Fig. \ref{fig:range_dev} shows the range deviation. The
errorbars here capture the minimum, median and maximum points. As we
can see, until the number of nodes reach $10000$, the $s$ estimated at
each node is exactly the same. Even when the number of nodes is equal to
$10000$, the difference of the maximum/minimum $s$ with the median $s$
is only $1$. We expect in the real world, the number of domain names will
be much bigger and the error will be even smaller.

\subsection{Performance of GR cPIR \label{sec:gentry_perform}}
\subsubsection{Storage Overhead}
With GR cPIR, all clients and servers need to synchronize a list of
prime numbers, the number of which decides the maximum number of
blocks a client can retrieve from the server. This can be easily
achieved by sharing a range of consecutive prime numbers.
Theoretically, there is no limit to the number of shared primes.
But, in practice, since $m=1024$ is good enough for hiding clients'
target, storing one thousand prime numbers are enough. Therefore,
the number of shared primes would not pose a serious overhead for
the client and the server.

\subsubsection{Communication Overhead}
The advantage of GR cPIR is its near-optimal
communication overhead. Specifically, the communication from the
client to the server includes one index indicating the ending prime
number in the shared list, the generator of a cyclic group and a
composite modulus. The communication from the server to the client is
just a group element. For example, if we set the modulus to be $2048$
bits, we can safely retrieve a block of length $512$ bits from a
database of $n$ blocks. Then we have the generator, the modulus and
the prime index to be $2048$ bits, $2048$ bits and $\log n$ bits
respectively.  The response from the server is $2048$ bits. Therefore
the total communication overhead is $\log n + 6144$ bits. If the block
length $l > 512$ bits, we can split each block into chunks with 512
bits each and execute $\lceil \frac{l}{512} \rceil$ instances of cPIR
in parallel. Since these instances share the same generator and
modulus, the client only needs to send the generator and modulus
once. And the communication overhead becomes $\log n + 4096 +
2048*\lceil \frac{l}{512} \rceil$ bits. This is an tremendous
improvement over the communication, compared to trivially sending all
the blocks whose communication is $nl$
bits. Fig. \ref{fig:gentry_compression} shows the compression ratio
under different combination of $l$ and $n$. As we can see, the larger
$l$ or $n$ is, the higher the compression ratio is.

\subsubsection{Computation Overhead}
\begin{figure*}[!t]
\begin{minipage}[b]{0.33\linewidth} 
\centering
\includegraphics[width=5.3cm]{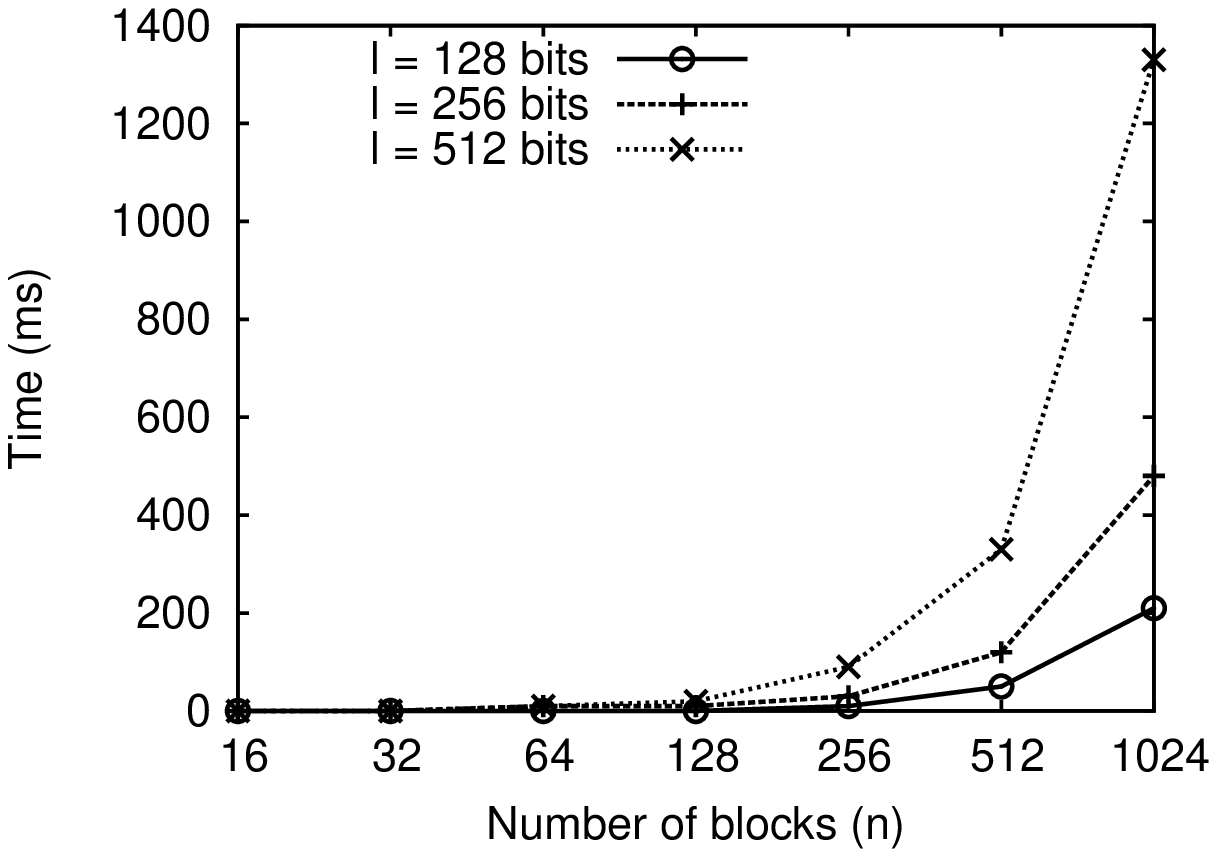}
\caption{Speed of Chinese Remainder Algorithm}
\label{fig:CRT_speed}
\end{minipage}
\hspace{0.1cm} 
\begin{minipage}[b]{0.33\linewidth}
\centering
\includegraphics[width=5.3cm]{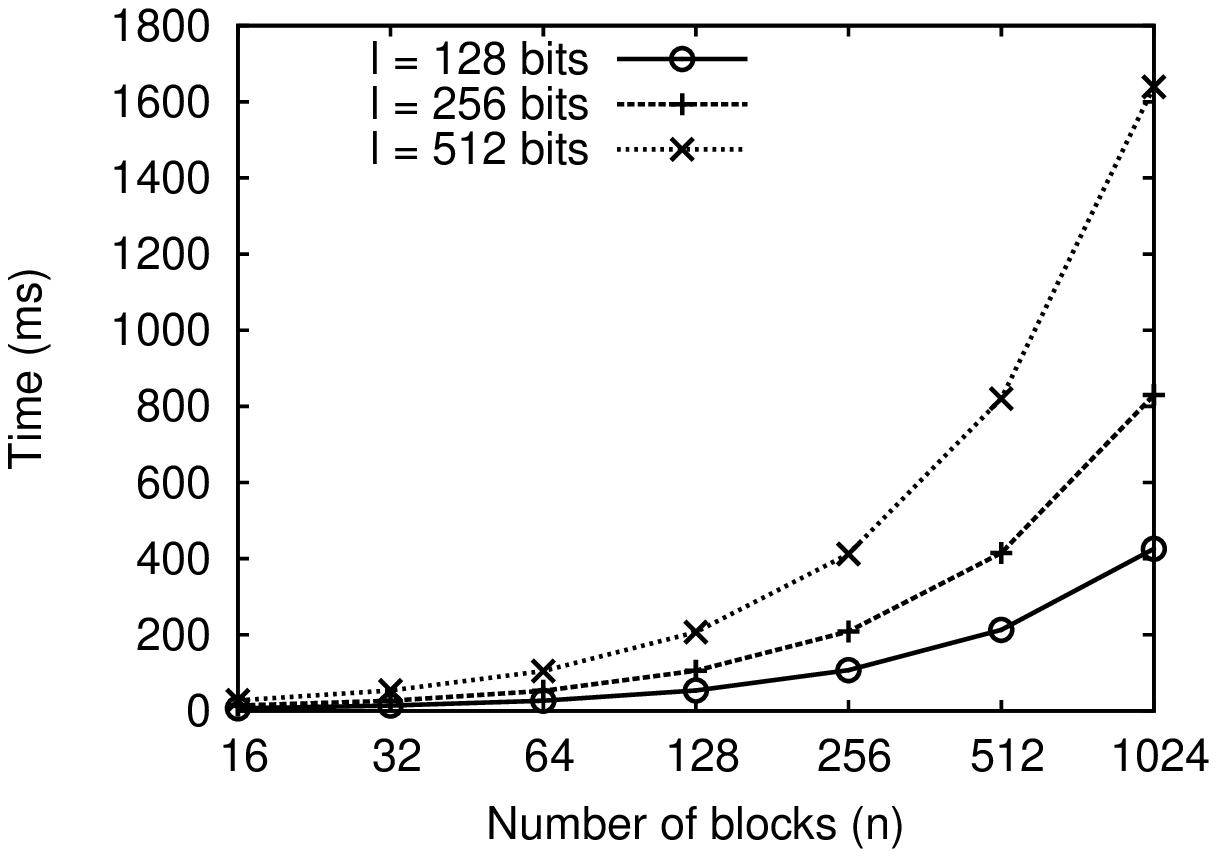}
\caption{Speed of modulus exponentiation}
\label{fig:exponentiation_speed}
\end{minipage}
\hspace{0.1cm} 
\begin{minipage}[b]{0.33\linewidth}
\centering
\includegraphics[width=5.3cm]{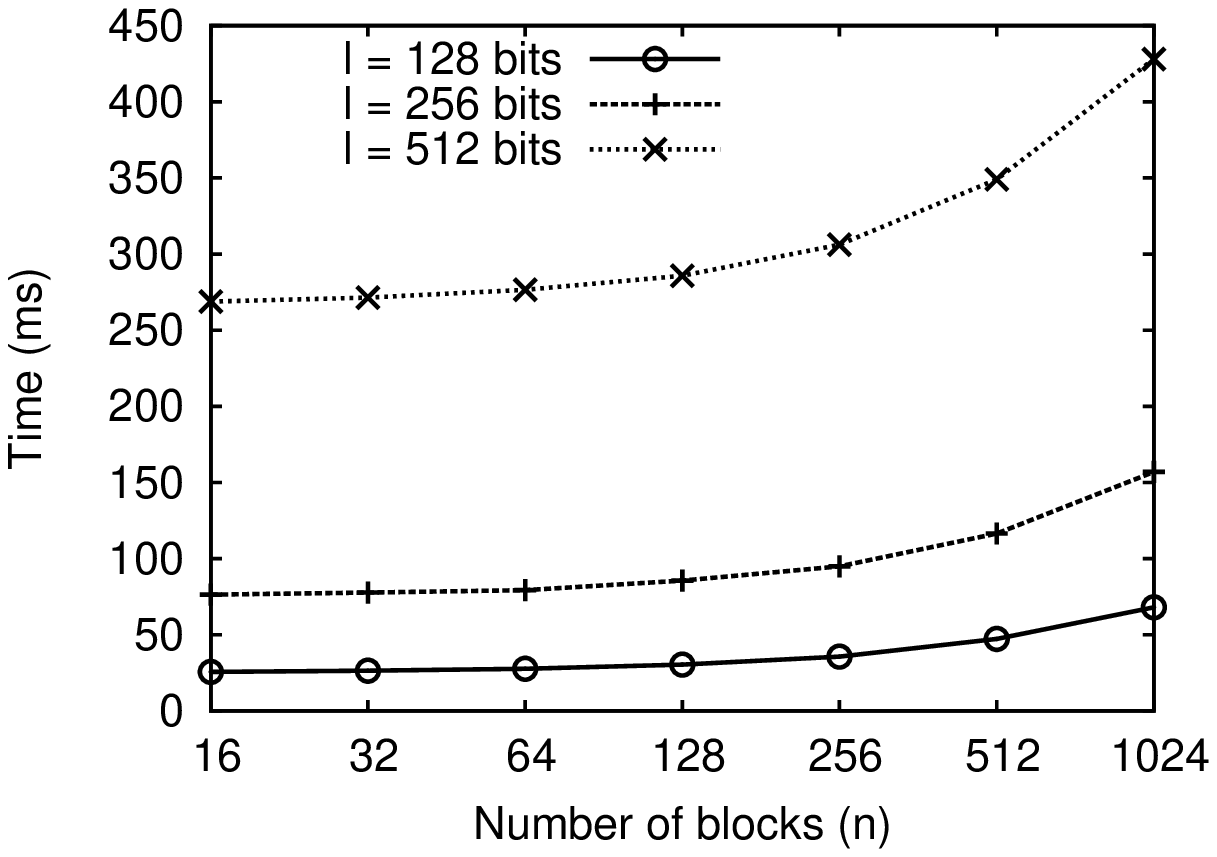}
\caption{Speed of Pohlig-Hellman algorithm}
\label{fig:Pohlig_speed}
\end{minipage}
\end{figure*}



We implemented GR cPIR by using GMP library~\cite{gmp} and tested it
on a desktop with Intel Core i7-920 (2.66MHz and 8M cache, MSRP:
$\$279$) inside. We choose the group modulus to be $2048$ bits long
which is expected to be safe until 2030 \cite{rsa_keysize}.  As
described in Sec.~\ref{gentry_pir}, GR cPIR includes three
operations: (1) server representing the whole database as an integer
$e$ through Chinese Remainder Theorem, (2) server responding to
client's query by modular exponentiation $g^e$. (3) client computing
discrete logarithm by Pohlig-Hellman algorithm.

Fig. \ref{fig:CRT_speed} shows the speed of computing congruence by
Chinese Remainder Theorem. We can see that computing $512$ bits
congruence takes negligible time until $n$ reaches $128$, where the
time spent for $l=512$ bits is $20$ ms, which means real time
congruence computation for $n \le 128$ is possible. We can also see
that the computation overhead increases super linearly with the
block size and with the number of retrieved elements $n$.  However,
a local PPDNS node does not always need to do this operation on
demand. If a local PPDNS server has a cache of the range of
responses, it can precompute $e$ for this range to save time for
future queries.

\begin{figure}[!t]
\centering
\includegraphics[width=2.5in]{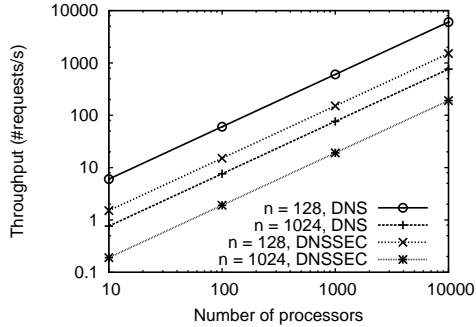}
\caption{Number of concurrent cPIR requests a server can support}
 \label{fig:numuser_processor}
\end{figure}

Fig. \ref{fig:exponentiation_speed} shows the computation time of
modular exponentiation $g^e$. Since $e$ is a representation of the
whole database, its size is up to $n \cdot l$ bits. From Fig.
\ref{fig:exponentiation_speed}, we can see that the speed of modular
exponentiation is still quite slow. To retrieve $32$ blocks of $512$
bits each, it takes $54$ms. The maximum size of a regular DNS
response is $512$ bytes ($2$k bytes for DNSSEC). In other words, to
retrieve a DNS response, the worst time is $432$ms for regular DNS
($1728$ms for DNSSEC). However, if we consider a server equipped
with multiple processors, each $512$-bit block retrieval can be
executed in parallel, therefore it is possible to make this scheme
scalable. These processors can be some cheap special-purpose
FPGAs~\cite{Michalski06} which can work faster than general-purpose
CPU in computing modular exponentiation. Fig.
\ref{fig:numuser_processor} shows the number of requests a server
can serve in one second with respect to the number of processors it
uses. From the figure, we can see that at least $100$ processors are
needed to serve $10\%$ concurrent requests if we assume the server
receives $1000$ requests per second. This actually matches Sion's
results~\cite{Sion07onthe} that it is computationally expensive to
deploy cPIR. However, as long as one is willing to invest money on
the server side, cPIR can still be practical in reducing
communication overhead.


Fig. \ref{fig:Pohlig_speed} shows the speed of extracting the exponent
from the response.
As we can see, the time increases quadratically with $l$ and is stable
for small $n$. Retrieving $512$-bit exponent takes up to $430$ms for
$n = 1024$. However, the waiting time occurs only at the client side
and it does not block other user's response. A user who executes cPIR
should be ready for some reasonable waiting time (several seconds).

In conclusion, GR cPIR works impressively well in reducing traffic between
the client and the server. However, it also poses high computation
overhead on both sides. Fortunately, we notice there are some
high-speed efficient cPIR variants~\cite{trostle07,
  Aguilar-melchor_alattice-based} and we expect more efficient cPIR schemes
to appear in the future. As PPDNS's support for cPIR is flexible, we
can accommodate any promising cPIR schemes into the system.

\subsection{Simulation Results \label{sec:simulation}}
\label{sec:tcp}
In this section, we examine PPDNS's performance through simulations
built on NS-3~\cite{ns3} and show that PPDNS can efficiently do target
lookup while incuring reasonable cost to the backbone network.

We compare the average delay, maximum link utilization and cache hit
ratio of PPDNS with standard single query and random-set query
performed in DHT-based DNS. We also study the range split ratio in
PPDNS.
\subsubsection{Setup\label{sec:setup}}
We collected the DNS trace at one of UCI's name servers between June
and July 2009. We exclude those non-legitimate names by querying
each distinct names in the trace, which results in $1,980,623$
distinct legitimate fully-qualified names. For each legitimate name,
its TTL and response size are recorded. We feed each simulator node
in NS-3 with one distinct day's DNS trace from 12:00 pm to 22:00 pm.
We choose Chord~\cite{Stoica2001Chord} as the underlying DHT for
PPDNS. Each simulator nodes' identifier is randomly picked from the
hash space.

\begin{table}
\centering
\begin{small}
\begin{tabular}{c|c|c|c}
Topology & Type & \# of Nodes & \# of Links \\
\hline
\hline
Abilene & R & 11 & 28 \\
Geant & R & 23 & 74 \\
Sprint & I & 52 & 168 \\ 
Tiscali & I & 41 & 174 \\ 
Brite-1 & A & 50 & 342 \\
Brite-2 & A & 50 & 386 \\
\hline
\end{tabular}
\end{small}
\caption{\label{tab:topo} \small Summary of the topologies used in our
simulation.  Types \textit{R}, \textit{I}, and \textit{A} mean \textit{Real},
\textit{Inferred}, and \textit{Random} topologies respectively.}
\end{table}

We simulate PPDNS with backbone topologies as we expect that PPDNS
should be deployed at domain level. We use real, inferred, and
synthetic topologies in our simulations. For real ones we use the
Abilene backbone topology and the Geant backbone topology. We also use
the topologies inferred by the Rocketfuel project~\cite{SpringMWA04}.
In addition, we generate synthetic topologies using the
BRITE~\cite{brite} topology generator. All real and inferred
topologies are on the level of Point-of-Presenses
(PoPs). Table~\ref{tab:topo} summarizes the topologies.

For real topologies, we use the actual link capacity in our study. For
Rocketfuel and random topologies, the link capacities are set
according to a two-tier model~\cite{kandula_walking:2005} where each
link is either OC48 (i.e., 2.48Gbps) or OC192 (i.e.,
10Gbps). Specifically, Links connecting to the top 20\% PoPs with the
highest node degrees have the higher capacity (10Gbps), while other
links have the lower capacity (2.48 Gbps). In addition, we make each
link's initial utilization to be at $60\%$ to simulate a median loaded
network. We set the propagation delay of the link with the
smallest weight to $10$ms and make all the other links' delays
proportional to their weights.

\subsubsection{Maximum link utilization}
\begin{figure*}[!t]
\begin{minipage}[b]{0.33\linewidth} 
\centering
\includegraphics[width=5.3cm]{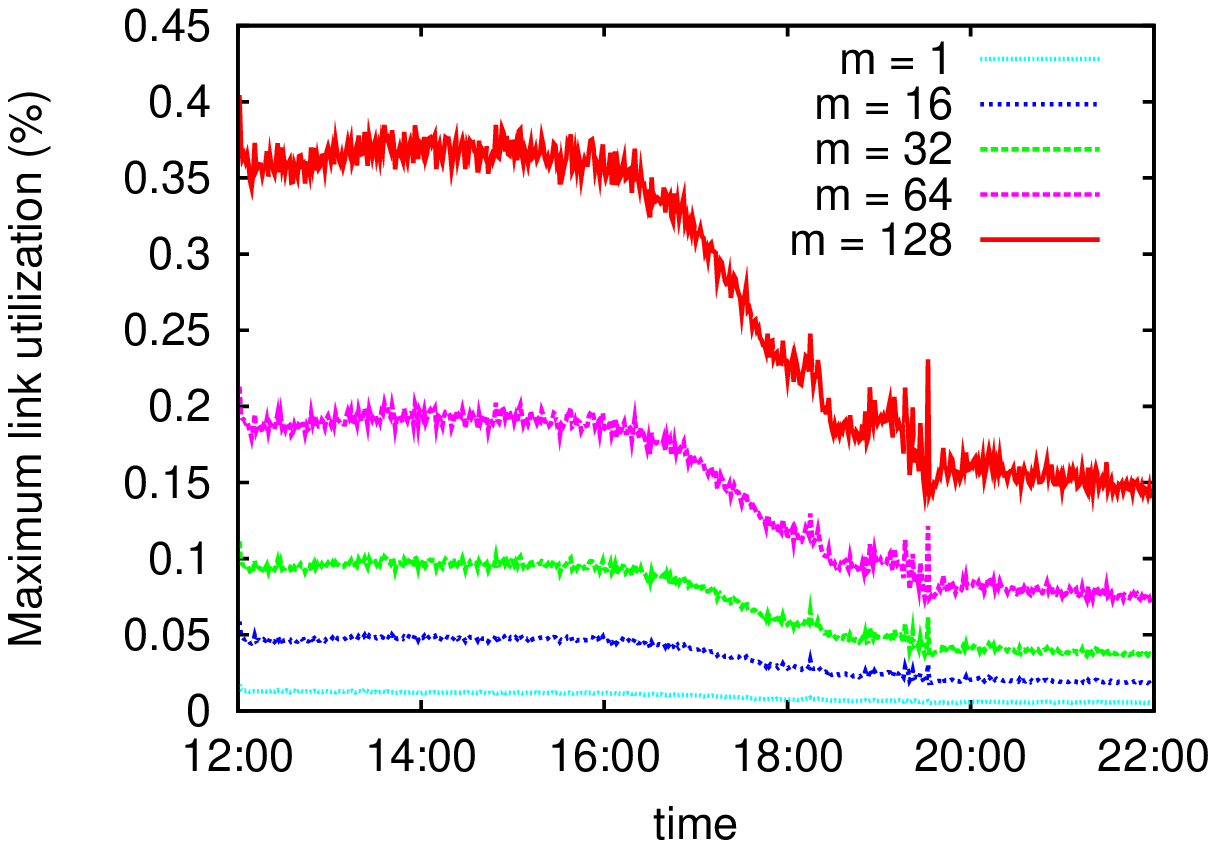}
\caption{Maximum link utilization for our range query}
\label{fig:50nodes_range_traffic}
\end{minipage}
\hspace{0.1cm} 
\begin{minipage}[b]{0.33\linewidth}
\centering
\includegraphics[width=5.3cm]{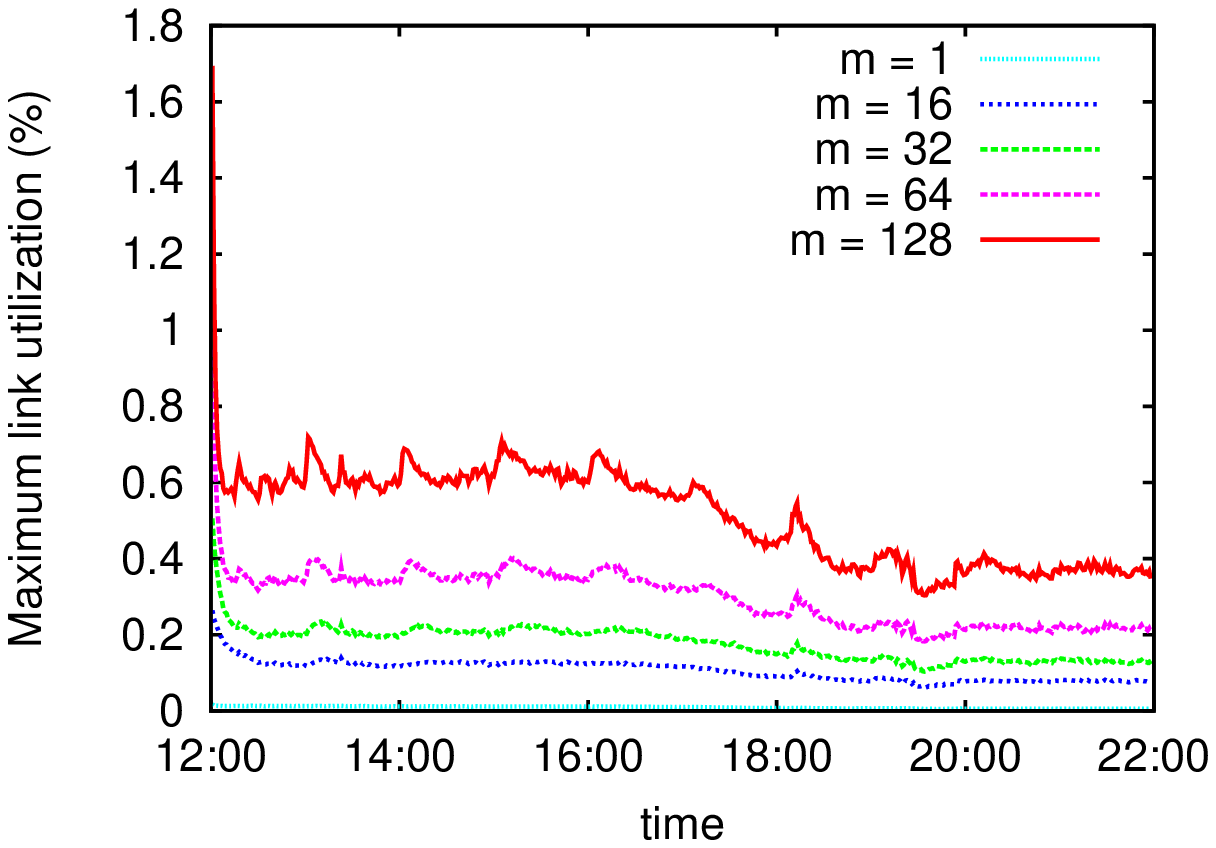}
\caption{Maximum link utilization for random-set query}
\label{fig:50nodes_random_traffic}
\end{minipage}
\hspace{0.1cm} 
\begin{minipage}[b]{0.33\linewidth}
\centering
\includegraphics[width=5.3cm]{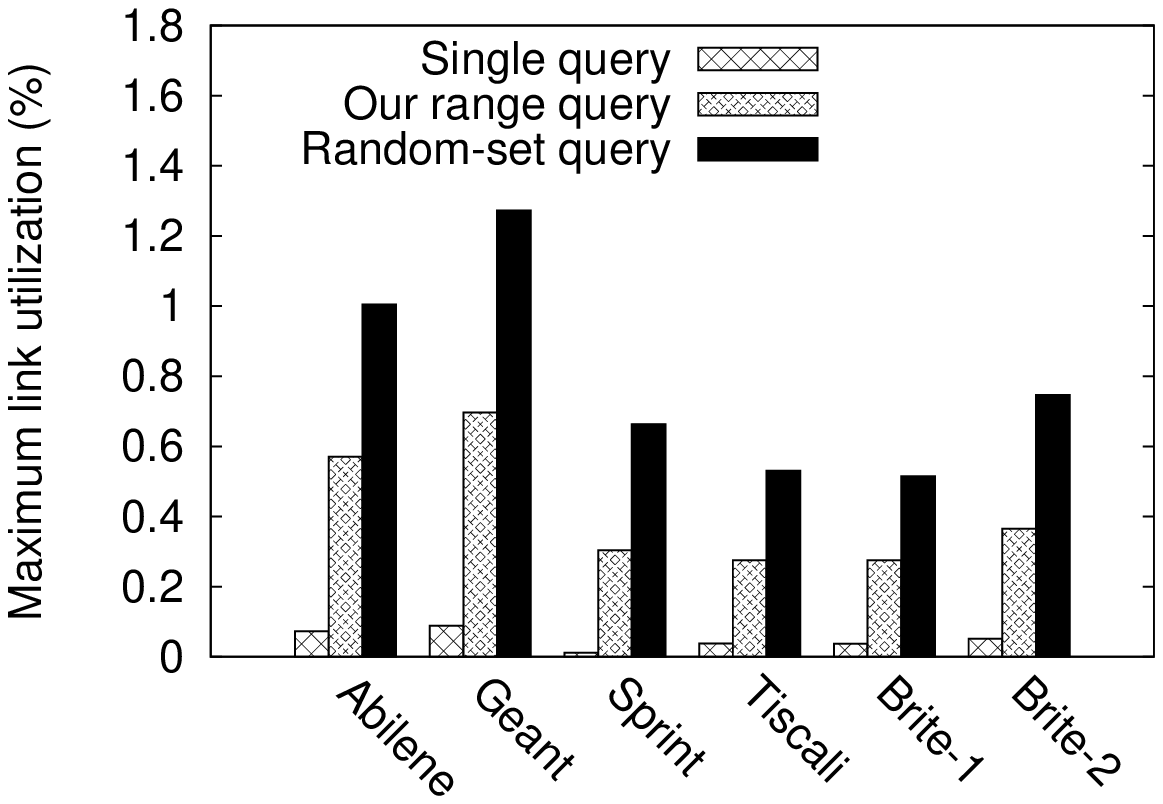}
\caption{Comparison of average maximum link utilization among all topologies}
\label{fig:alltopo_traffic}
\end{minipage}
\end{figure*}





Fig.~\ref{fig:50nodes_range_traffic} and
Fig.~\ref{fig:50nodes_random_traffic} compares the variation of the
maximum link utilization with time in the case of the random topology
for our range query and random-set query. The line marked with
$m=1$ means single query. We omit other topologies here because they
all show the same trend. The link utilization we show here excludes
the link's initial load and is purely contributed by DNS queries. As
we can see, the maximum link utilization almost doubles as the range
size doubles. Our range-query scheme sees less maximum link utilization
than random-set query.

To further compare the link utilization between our range query and
random-set query, Fig.~\ref{fig:alltopo_traffic} shows the maximum
link utilization averaged over time for different topologies where $m$
is fixed at $128$. We can see that our range query is much better than
random query in terms of maximum link utilization. This is due to the
fact that the local PPDNS node has a high chance of caching the whole
range if there is a previous range query to any target inside this
range.  While as to random-set query, even if the same target has been
queried before, there is little chance PPDNS can cache all responses
for every name inside the random set of a new query for the
same target since the random set chosen each time is random.

\subsubsection{Average Delay}
\begin{figure*}[!t]
\begin{minipage}[b]{0.33\linewidth} 
\centering
\includegraphics[width=5.3cm]{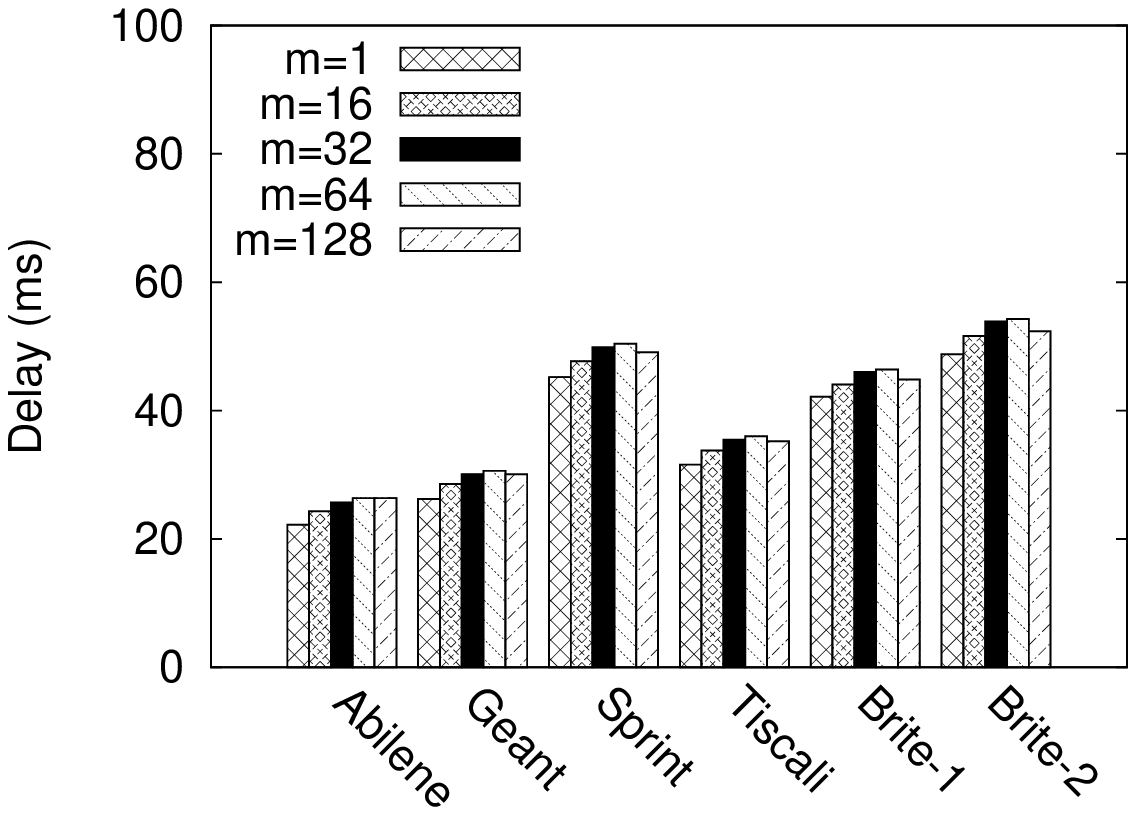}
\caption{Average query delay for our range query}
\label{fig:range_delay}
\end{minipage}
\hspace{0.1cm} 
\begin{minipage}[b]{0.33\linewidth}
\centering
\includegraphics[width=5.3cm]{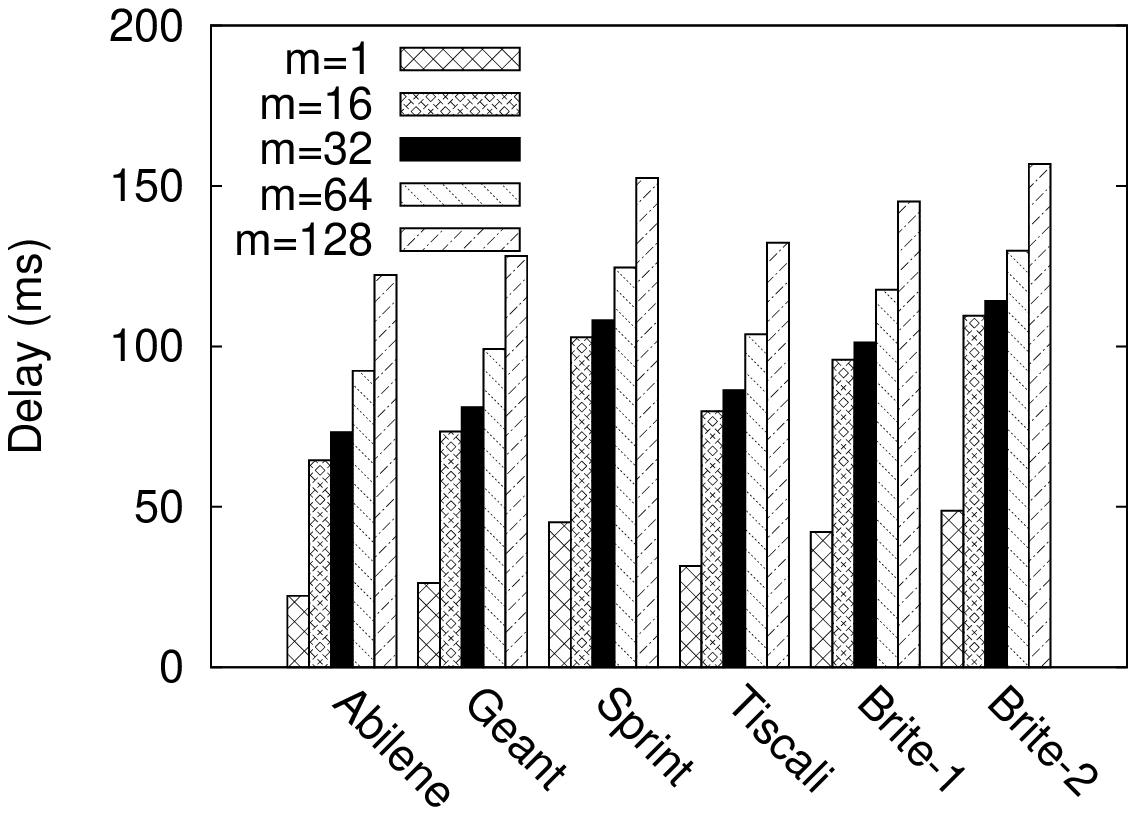}
\caption{Average query delay for random-set query}
\label{fig:random_delay}
\end{minipage}
\hspace{0.1cm} 
\begin{minipage}[b]{0.33\linewidth}
\centering
\includegraphics[width=5.3cm]{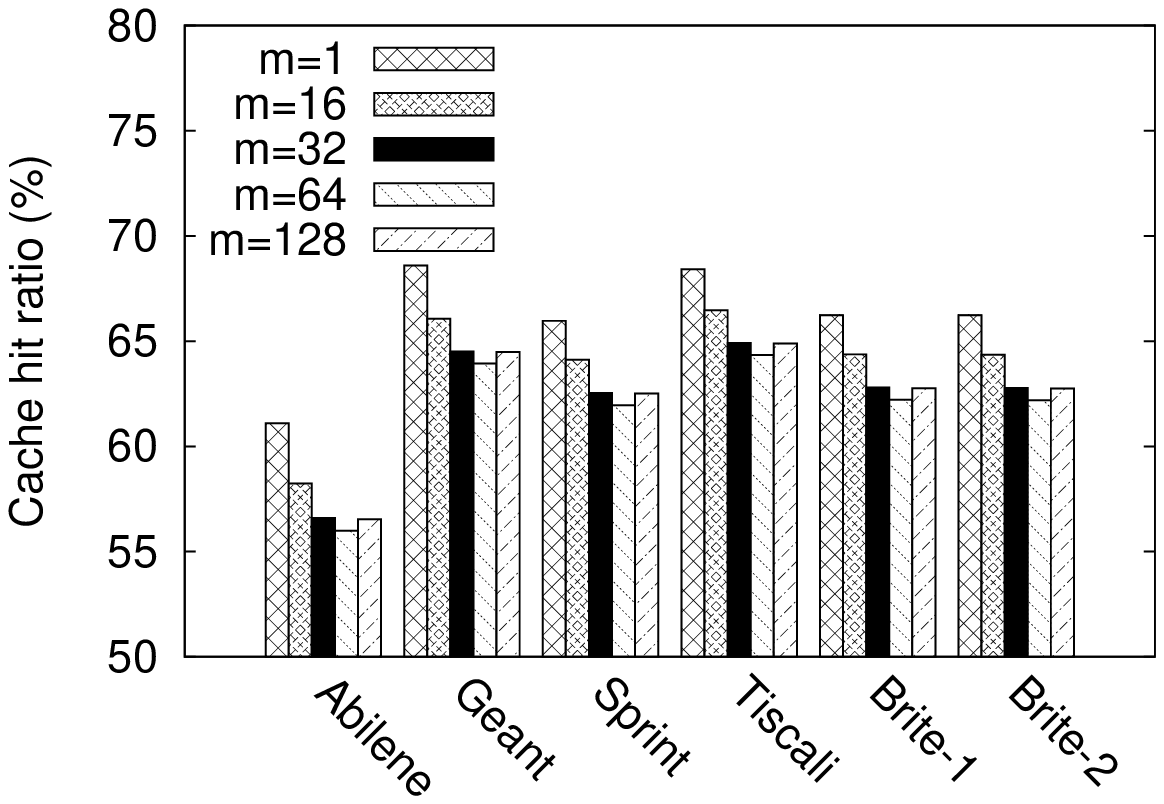}
\caption{Cache hit ratio for our range query}
\label{fig:cache_hit}
\end{minipage}
\end{figure*}

The delay calculated here is caused purely by backbone network, not
including the delay between end host and local PPDNS node.
Fig.~\ref{fig:range_delay} shows the average delay for our range
query. As we can see, the average delay in different cases is almost
the same as a single query (e.g. $m=1$). This again is due to our
range query's better capability of being cached. We measure the delay
for a random set query as the delay between when the first request is
sent and when the last response is received for one query set. From
Fig.~\ref{fig:random_delay}, we can see the average delay for
random-set query, is much higher than that for the single query and
our range query.

\subsubsection{Cache hit ratio}


Fig. \ref{fig:cache_hit} shows the cache hit ratio for our range query. We can see that
the cache hit ratio decreases first as query range size increases and
then the ratio increases. This is because the TTL for a range is
decided by the smallest TTL of all domain names' TTL inside the
range. By increasing the range, the smallest TTL has a chance to
decrease and therefore the hit ratio decreases as well. However,
further increasing range size also helps cache more results which
increases the probability of the range being hit. That is why cache hit ratio at
$m=128$ is higher than that at $m=64$. Also we can see that the cache
hit ratio is over $50\%$ for our range query in all cases. In
contrast, the cache hit ratio for random-set query is zero throughout
our study and we do not show it here.

\subsubsection{Range split ratio}
\begin{figure}[!t]
\centering
\includegraphics[width=5.3cm]{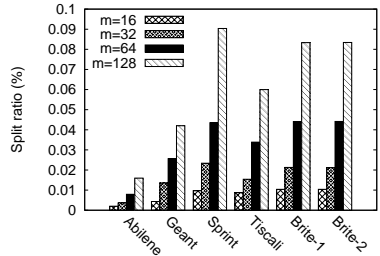}
\caption{Range split ratio for our range query}
\label{fig:split_ratio}
\end{figure}
Finally, we measure the split ratio for the range query. Range split
can cause extra response delay, since local PPDNS needs to wait
for all split-ranges' responses to arrive before answering its
client. Fig.~\ref{fig:split_ratio} shows the percentage of queries which are split in our
study. In all cases we examined, the split ratio is below $0.1\%$. And it
is linearly proportional to query range size. We only use $1,980,623$
legitimate domain names in our simulation.  If we increase the number of available
domain names (up the density), the split ratio should decrease
accordingly. In practice, we expect the total number of legitimate
domain names to be 100 times larger than we recorded. Even with a
range size of $10,000$, the split ratio should remain below $0.1\%$.

\section{Conclusion\label{sec:conclusion}}
In this paper, we propose a new Privacy-Preserving Domain Name
System, called PPDNS. PPDNS is built upon DHT and employs a special
range query which takes advantage of DHT index structure to yield a
secure system. Meanwhile PPDNS incorporates an adjustable cPIR
framework which can help clients reduce their communication
overhead.

Security analysis shows that PPDNS significantly improves privacy
for name queries. Performance evaluation demonstrates that PPDNS
incurs reasonable communication overhead to the backbone links. It
also shows that PPDNS retains the efficient part of the traditional
DNS--high cache hit ratio, and incurs much lower latency than
random-set query.


\vspace{-0.1in}

\footnotesize
\bibliographystyle{abbrv}
\bibliography{yanbinl}

\end{document}